\documentclass[fleqn,usenatbib]{mnras}

\usepackage{newtxtext,newtxmath}
\usepackage[T1]{fontenc}
\usepackage{arydshln}
\DeclareRobustCommand{\VAN}[3]{#2}
\let\VANthebibliography\thebibliography
\def\thebibliography{\DeclareRobustCommand{\VAN}[3]{##3}\VANthebibliography}

\usepackage{graphicx}	% Including figure files
\usepackage{amsmath}	% Advanced maths commands
\usepackage{float}
\usepackage[autopunct=true]{csquotes}

\newcommand{\acknowledgments}[1]{\begin{small}\section*{Acknowledgments}\end{small}{\noindent #1}\vspace{5pt}}
\newcommand{\datastatement}[1]{\begin{small}\section*{Data Availability Statement}\end{small}{\noindent #1}\vspace{5pt}}

\newcommand{\Bangle}{\theta_{B}}
\newcommand{\Alf}{{Alfv\'en}}

\newcommand{\fref}[1]{Fig.~\ref{#1}}

\newcommand{\Dt}[1]{\frac{\mathrm{d} #1}{\mathrm{dt}}}
\newcommand{\initvalupper}[1]{#1^{0}}
\newcommand{\initvallower}[1]{#1_{0}}
\newcommand{\driftvel}{{\bf w}_{s}}%{{\boldsymbol{\sizeparam}}}
\newcommand{\driftvelhat}{\hat{{\bf w}}_{s}}%{\hat{\boldsymbol{\sizeparam}}}
\newcommand{\dustvel}{{\bf v}_{d}}
\newcommand{\gasvel}{{\bf u}_{g}}
\newcommand{\gasden}{\rho_{g}}
\newcommand{\rhobase}{\rho_{\rm base}}

\newcommand{\dustden}{\rho_{d}}

\newcommand{\rhogas}{\gasden}

\newcommand{\ts}{t_{s}}
\newcommand{\cs}{c_{s}}
\newcommand{\vA}{v_{A}}
\newcommand{\tL}{t_{L}}
\newcommand{\grainsuff}{_{\rm grain}}
\newcommand{\internaldensity}{\bar{\rho}\grainsuff^{\,i}}
\newcommand{\grainsize}{\epsilon\grainsuff}

\newcommand{\grainmass}{m\grainsuff}
\newcommand{\graincharge}{q\grainsuff}
\newcommand{\grainchargeZ}{Z\grainsuff}
\newcommand{\grainsizemax}{\grainsize^{\rm max}}

\newcommand{\B}{{\bf B}}

\newcommand{\Bhat}{\hat\B}

\newcommand{\Lbox}{L_{\rm box}}
\newcommand{\Lscale}{H_{\rm gas}}
\newcommand{\sizeparam}{\tilde{\alpha}}
\newcommand{\sizeparammax}{\sizeparam_{\rm m}}
\newcommand{\chargeparam}{\tilde{\phi}}
\newcommand{\chargeparammax}{\chargeparam_{\rm m}}
\newcommand{\accparam}{\tilde{a}_{\rm d}}

\newcommand{\gravparam}{\tilde{g}}
\newcommand{\dustgas}{\mu^{\rm dg}}

\def\app#1#2{%
  \mathrel{%
    \setbox0=\hbox{$#1\sim$}%
    \setbox2=\hbox{%
      \rlap{\hbox{$#1\propto$}}%
      \lower1.1\ht0\box0%
    }%
    \raise0.25\ht2\box2%
  }%
}

\title[Dust Dynamics in AGN Winds]{Dust Dynamics in AGN Winds: A New Mechanism For Multiwavelength AGN Variability}

\author[Soliman et al.]{
\parbox[t]{\textwidth}{
Nadine H.~Soliman$^1$,
Philip F.~Hopkins$^1$
} \vspace*{4pt} \\
$^1$ TAPIR, Mailcode 350-17, California Institute of Technology, Pasadena, CA 91125, USA \\
}

\date{Accepted XXX. Received YYY; in original form ZZZ}

\pubyear{2021}

\begin{document}
\label{firstpage}
\pagerange{\pageref{firstpage}--\pageref{lastpage}}
\maketitle

\begin{abstract}

Partial dust obscuration in active galactic nuclei (AGN) has been proposed as a potential explanation for some cases of AGN variability. The dust-gas mixture present in AGN tori is accelerated by radiation pressure, leading to the launching of an AGN wind. Dust under these conditions has been shown to be unstable to a generic class of fast-growing resonant drag instabilities (RDIs). In this work, we present the first numerical simulations of radiation-driven outflows that explicitly include dust dynamics in conditions resembling AGN winds. We investigate the implications of RDIs on the torus morphology, AGN variability, and the ability of radiation to effectively launch a wind. We find that the RDIs rapidly develop, reaching saturation at times much shorter than the global timescales of the outflows, resulting in the formation of filamentary structure on box-size scales with strong dust clumping and super-\Alf{ic} velocity dispersions. The instabilities lead to fluctuations in dust opacity and gas column density of 10-20\% when integrated along mock observed lines-of-sight to the quasar accretion disk. These fluctuations occur over year to decade timescales and exhibit a red-noise power spectrum commonly observed for AGN. Additionally, we find that the radiation effectively couples with the dust-gas mixture, launching highly supersonic winds that entrain 70-90\% of the gas, with a factor of $\lesssim 3$ photon momentum loss relative to the predicted multiple-scattering momentum loading rate. Therefore, our findings suggest that RDIs play an important role in driving the clumpy nature of AGN tori and generating AGN variability consistent with observations.
\end{abstract}


%%%%%%%%%%%%%%%%%%%%%%%%%%%%%%%%%%%%%%%%%%%%%%%%%%

%%%%%%%%%%%%%%%%% BODY OF PAPER %%%%%%%%%%%%%%%%%%

\section{Introduction} \label{sec:intro}
Dust plays a critical role in how a wide range of astrophysical systems form, evolve, and are observed. It is involved in processes such as planetary formation and evolution \citep{lissauer1993planet, liu2020tale,apai:dust.review}; chemical evolution \citep{watanabe2008ice,whittet1993dust, weingartner2001electron, minissale2016dust}, heating, and cooling within the interstellar medium (ISM) and star formation \citep{dorschner:dust.mineralogy.review,weingartner:2001.grain.charging.photoelectric,draine:2003.dust.review, salpeter1977formation, spitzer2008physical};  as well as feedback and outflow launching in star-forming regions, cool stars and active galactic nuclei (AGN) \citep{king2015powerful, murray2005maximum,2018A&ARv..26....1H}. Moreover, dust imprints ubiquitous observable signatures, such as the attenuation and extinction of observed light \citep{savage1979observed,1984drainelee, mathis1990interstellar}. 

One particular regime where dust is believed to play a central role in both dynamics and observations is the ``dusty torus'' region around AGN \citep{antonucci:1982.torus,lawrence:1982.torus.alignment,urry1995unified, choi2022physical}. It is well established that outside of the dust sublimation radius, AGN and quasars are surrounded by a dust-laden region with extinction and column densities ranging from $\sim 10^{22}\,{\rm cm^{-2}}$ in the polar direction to $\sim 10^{26}\,{\rm cm^{-2}}$ in the mid-plane (on average), exhibiting  ``clumpy'' sub-structure in both dust and gas, ubiquitous time variability on $\gtrsim\,$yr timescales, and a diverse array of detailed geometric and reddening properties  (see \citealt{krolik1988molecular, elitzur:torus.wind, tristram2007resolving,nenkova2008agn, nenkova2008agnA,Stalevski:2012.dust.props.agn.var, leighly2015variable}, or for recent reviews see \citealt{netzer2015revisiting,padovani:agn.review,hickox.alexander:2018.agn.review,balokovic:2018.torus.fits}), as well as a broad variety of different extinction curve shapes \citep{laor.draine:1993.agn.spectra.dust,hopkins:dust,maiolino.2004:qso.dust.sne,hatziminaoglou:2009.torus.properties.inferred.obs,gallerani:2010.qso.extinction,hoenig:clumpy.torus.modeling}. It has been recognized for decades that the torus represents one (of several) natural locations where bright AGN should drive outflows, and indeed many have gone so far as to propose the ``torus'' is, itself, an outflow \citep[see e.g.][]{sanders88:quasars,pier.krolik:1992.torus.rad.pressure,koniglkartje:disk.winds,elvis:outflow.model,elitzur:torus.wind}. Put simply, because the dust cross-section to radiation scattering and absorption is generally much larger than the Thompson cross section, which defines the Eddington limit, any AGN accreting at even modest fractions of Eddington should be able to unbind material via radiation pressure on dust, launching strong outflows. This concept has led to an enormous body of detailed observational followup  \citep{hoenig:clumpy.torus.modeling,2008A&A...479..389H,2009A&A...502...67T,2009ApJ...695..781B,2011ApJ...736...82A,2011A&A...527A.121K,ricci:2017.torus.rad.pressure.mods.review,hoenig:2019.ir.submm.torus.wind.review} and detailed theoretical simulations and models of dust-radiation pressure-driven outflows from AGN in the torus region \citep{thompson:rad.pressure,debuhr:momentum.feedback,wada:torus.mol.gas.hydro.sims,wada:2012.torus.fountain,roth:2012.rad.transfer.agn,costa:dusty.wind.driving,thompson2015dynamics,ishibashi2015agn,chan:2016.rad.pressure.outflows.torus,baskin:2018.dust.inflated.acc.disks.torus,2018MNRAS.476..512I,kawakatu:2020.obscuration.torus.from.stellar.fb.in.torus, venanzi2020role}. 

Yet despite this extensive literature, almost all the theoretical work discussed above has assumed that the dust dynamics are perfectly coupled to the dynamics of the surrounding gas -- effectively that the two ``move together'' and the dust (even as it is created or destroyed) can simply be treated as some ``additional opacity'' of the gas. But in reality, radiation absorbed/scattered by grains accelerates those grains, which then interact with gas via a combination of electromagnetic (Lorentz, Coulomb) and collisional (drag) forces, re-distributing that momentum.

Accurately accounting for these interactions is crucial for understanding any radiation-dust-driven outflows. If the dust ``free-streaming length'' is very large, grains could simply be expelled before sharing their momentum with gas \citep{elvis:2002.qso.dust.production}. If dust can be pushed into channels, creating low-opacity sight-lines through which radiation can leak out efficiently, some authors have argued that the coupled photon momentum might be far smaller than the standard expectation $\sim \tau_{\rm IR}\,L/c$ (where $\tau_{\rm IR}$ is the infrared optical depth; see \citealt{krumholz:2012.rad.pressure.rt.instab} but also \citealt{kuiper:2012.rad.pressure.outflow.vs.rt.method,wise:2012.rad.pressure.effects,tsang:monte.carlo.rhd.dusty.wind}).

Perhaps most importantly, \citet{squire.hopkins:RDI} showed that radiation-dust-driven outflows are generically unstable to a class of ``resonant drag instabilities'' (RDIs). RDIs occur due to differences in the forces acting on the dust versus the gas and are inherently unstable across a broad range of wavelengths. However, the fastest growing modes,  ``resonant modes'', arise when the natural frequency of a dust mode matches that of a gas mode. Each pair of resonant modes leads to a unique instability with a characteristic growth rate, resonance and mode structure. In subsequent work \citep{hopkins:2017.acoustic.RDI,squire:rdi.ppd,hopkins:2018.mhd.rdi}, the authors showed that systems like radiation-dust-driven outflows are unstable to the RDIs on all wavelengths -- even scales much larger than the dust free-streaming length or mean free path. Subsequent idealized simulations of these instabilities \citep{moseley:2018.acoustic.rdi.sims,seligman:2018.mhd.rdi.sims,hopkins:2018.mhd.rdi} have shown that they can grow rapidly, reaching significant non-linear amplitudes on large scales. Furthermore, the simulations demonstrated time-dependent clustering in both dust and gas, and a separation of dust and gas that is dependent on grain size. Additionally, the RDIs could drive fluctuations in the local dust-to-gas ratios which would affect the absorption and re-emission of radiation at different wavelengths. Specifically, as dust dominates the variability in the optical-UV bands but has a weaker effect on the IR and X-ray bands, dust-to-gas fluctuations can result in differences in the observed variability of the AGN emission across the electromagnetic spectrum.

The insights gained from these simulations are crucial not only for determining the initiation of an outflow but also for explaining various related phenomena. These include clumping in the torus, variations in AGN extinction curves, and specific forms of temporal variability. AGN sources are known to exhibit variability at essentially all wavelengths and timescales, ranging from hours to billions of years \citep{uttley2004brief, paolillo2004prevalence, paolillo2017tracing, assef2018luminous, caplar2017optical}. However, there have been observations of sources where the X-ray flux varies by approximately 20\% to 80\% over a few years, with no apparent variation in the optical component \citep{2002ApJ...571..234R,2005ApJ...623L..93R, markowitz2014first, laha2020variable, de2007broad, smith2007x}. 
In some cases, 'changing-look' AGN have shown order of magnitude variability on timescales as short as a few days to a couple hours \citep[e.g.,][]{lamassa2015discovery, runnoe2016now, ruan2016toward, mcelroy2016close, yang2018discovery, mathur2018changing, wang2018identification, stern2018mid, ross2020first, trakhtenbrot20191es, hon2020changing}. However, the processes driving such variability and the clumpy nature of the torus remain unexplained. 

In this study, we investigate the behaviour of radiation-dust-driven outflows for AGN tori, including explicit dust-gas radiation dynamics for the first time. We introduce our numerical methods and initial conditions in \S   \ref{sec:methods}, followed by an analysis of our results in \S \ref{sec:results}. We analyze the morphology, dynamics, and non-linear evolution of the dusty gas in the simulations, and in \S \ref{subsec:radhydro} we compare our standard simulations results to simulations with full radiation-dust-magnetohydrodynamics. Additionally, we investigate the feasibility of launching radiation-driven outflows and measure the momentum coupling efficiency within the wind in  \S \ref{subsec:launch}. In  \S \ref{subsec:var}, we examine how the presence of RDIs affects observable AGN properties, such as time variability. Finally, we provide a summary of our findings in  \S \ref{sec:conc}.

\section{Methods \&\ Parameters}
\label{sec:methods}

\begin{figure*}
    \centering
    \includegraphics[width=0.95\textwidth]{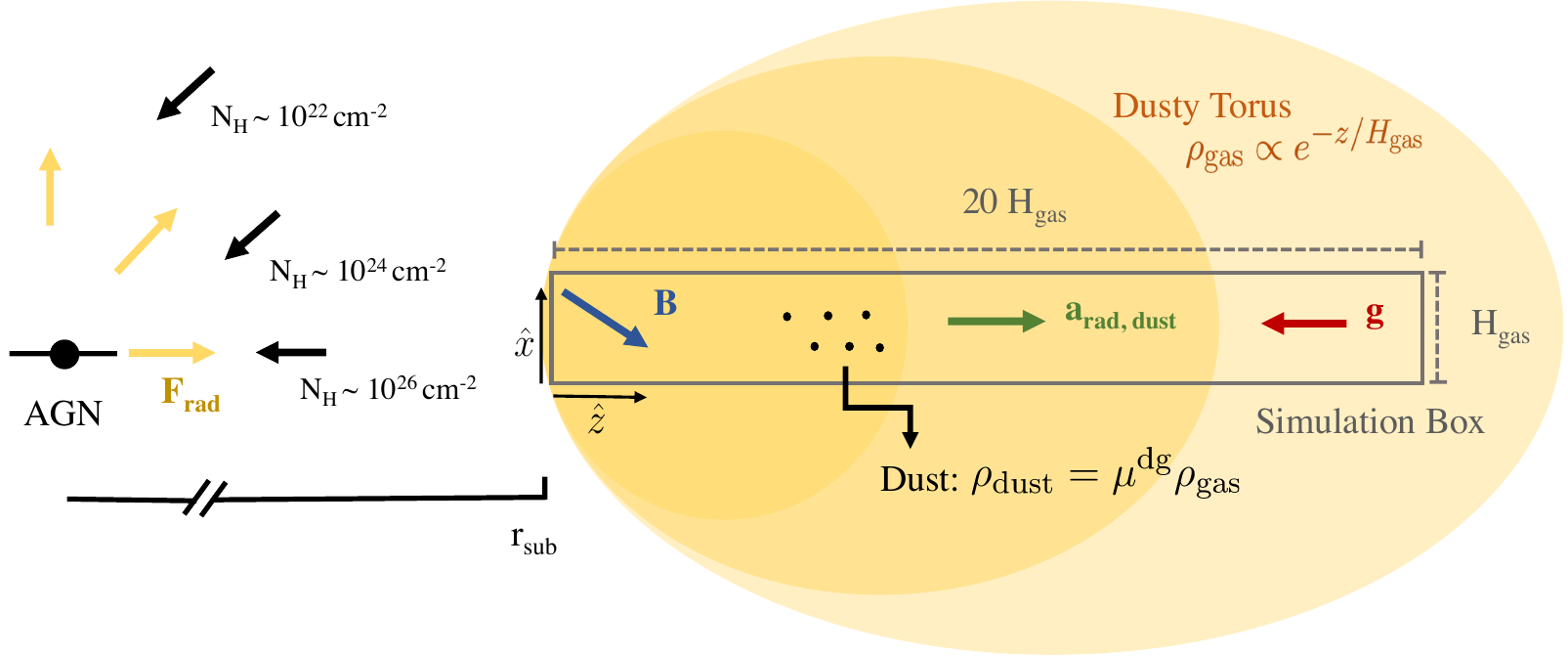}
    \caption{Cartoon illustrating our simulation setup. We simulate 3D boxes of size $\Lscale \times \Lscale \times \,  20 \, \Lscale$ along the $\hat{x}$, $\hat{y}$ and $\hat{z}$ directions respectively with $\sim 10^6$ resolution elements. We enforce outflow upper and reflecting lower boundary conditions with periodic sides. The gas and dust are initially stratified such that $\rho_{\rm gas} \propto e^{-z/\Lscale}$, and $\rho_{\rm d} = \dustgas_{0} \rho_{\rm gas}$ where $\dustgas_{0}=0.01$ corresponding to a uniform dust-to-gas ratio. The gas follows an isothermal ($\gamma = 1$) EOS with sound speed $c_s$, an initial magnetic field $\bf{B_0} = |\bf{B}| (\rm{sin \theta^0_B} \hat x + \rm{cos \theta^0_B} \hat z)$ in the $\hat{x} - \hat{z}$ plane and gravitational acceleration $\bf{g}$ $= -g \hat{z}$. The dust grains are modelled as super-particles each representing a population of grains of a given size sampled from a standard MRN spectrum with a factor = 100 range of sizes. The grains are photo-electrically charged, with the charge appropriately scaled according to grain size. They experience an upward acceleration $\rm a_{rad, dust}$ due to absorption of an initial upward radiation flux $F_0 = + F_0 \hat{z}$ corresponding to radiation from an AGN located a sublimation radius $\rm r_{sub}$ distance away, and are coupled to the gas through drag and Lorentz forces. We consider a range of $10^{22} - 10^{26} \, \rm cm^{-2}$ in column densities representing different lines-of-sight angles through the dusty torus.}
    \label{fig:cartoon}
\end{figure*}

We consider an initially vertically-stratified mixture of magnetized gas (obeying the ideal MHD equations) and an observationally-motivated spectrum of dust grains with varying size, mass, and charge. The dust and gas are coupled to one another via a combination of electromagnetic and collisional/drag forces. The system is subject to an external gravitational field, and the dust absorbs and scatters radiation from an external source. In Figure \ref{fig:cartoon}, we show a cartoon illustrating the geometry of our idealized setup and its relation to an AGN torus.

\subsection{Numerical Methods}

The numerical methods for our simulations are identical to those in \citet{hopkins:2021.dusty.winds.gmcs.rdis}, to which we refer for more details (see also \citealt{hopkins.2016:dust.gas.molecular.cloud.dynamics.sims,lee:dynamics.charged.dust.gmcs,moseley:2018.acoustic.rdi.sims,seligman,hopkins:2019.mhd.rdi.periodic.box.sims,steinwandel:2021.dust.rdi.variable.stars,ji:2021.cr.mhd.pic.dust.sims,squire:2022.acoustic.rdi.size.spectrum} for additional details and applications of these methods). Briefly, we run our simulations with the code {\small GIZMO}\footnote{A public version of the code is available at \href{http://www.tapir.caltech.edu/~phopkins/Site/GIZMO.html}{\url{http://www.tapir.caltech.edu/~phopkins/Site/GIZMO.html}}} \citep{hopkins:gizmo}, utilizing the Lagrangian ``meshless finite mass method'' (MFM) to solve the equations of ideal magnetohydrodynamics (MHD; \citealt{hopkins:mhd.gizmo,hopkins:cg.mhd.gizmo,hopkins:gizmo.diffusion,su:2016.weak.mhd.cond.visc.turbdiff.fx}). Dust grains are modelled as ``super-particles'' \citep{carballido:2008.grain.streaming.instab.sims,johansen:2009.particle.clumping.metallicity.dependence,bai:2010.grain.streaming.vs.diskparams,pan:2011.grain.clustering.midstokes.sims,2018MNRAS.478.2851M} where each simulated ``dust particle'' represents an ensemble of dust grains with a similar grain size $(\epsilon_{\text{grain}})$, charge $(q_{\text{grain}})$, and mass $(m_{\text{grain}})$. 

We simulate a 3D box with a base of length $\Lscale = L_{\rm xy}$ in the $xy$ plane and periodic $\hat{x}$, $\hat{y}$ boundaries, and height $\Lbox = L_{\rm z}=20\,L_{\rm xy}$ in the $\hat{z}$ direction with a reflecting lower ($z=0$) and outflow upper ($z=+L_{\rm z}$) boundary. Dust and gas feel a uniform external gravitational field ${\bf g} = -g\,\hat{z}$. The gas has initial uniform velocity  $\initvalupper{\gasvel} = 0$, initial magnetic field $\initvallower{\B} \equiv \initvallower{B}\,\initvallower{\Bhat}$ in the $xz$ plane ($\initvallower{\Bhat} = \sin({\Bangle^{0}})\,\hat{x} + \cos({\Bangle^{0}})\,\hat{z}$), obeys a strictly isothermal equation of state ($P = \gasden\,\cs^{2}$), and the initial gas density is stratified with $\initvalupper{\gasden} \equiv \gasden(t=0) =\rhobase\,\exp{(-z/\Lscale)}$ (with $\rhobase \approx M_{\rm gas,\,box}/\Lscale^{3}$).

Each dust grain obeys an equation of motion
\begin{align}\label{eq:eom}
\Dt{\dustvel} &={\bf a}_{\rm gas,\,dust} + {\bf a}_{\rm grav} + {\bf a}_{\rm rad} \\
\nonumber &=  -\frac{\driftvel}{\ts} - \frac{\driftvel\times\Bhat}{\tL} + {\bf g} + \frac{\pi\,\grainsize^{2}}{m_{\rm grain}\,c}\,\langle Q \rangle_{\rm ext}\,{\bf G}_{\rm rad}
\end{align}
where $\dustvel$ is the grain velocity; $\driftvel \equiv \dustvel - \gasvel$ is the drift velocity for a dust grain with velocity $\dustvel$ and gas velocity $\gasvel$ at the same position ${\bf x}$; ${\bf B}$ is the local magnetic field; ${\bf a}_{\rm gas,\,dust} = -{\driftvel}/{\ts} - {\driftvel\times\Bhat}/{\tL}$ includes the forces from gas on dust including drag (in terms of the ``stopping time'' $\ts$) and Lorentz forces (with gyro/Larmor time $\tL$); ${\bf a}_{\rm grav} = {\bf g}$ is the external gravitational force; and ${\bf a}_{\rm rad}$ is the force from radiation in terms of the grain size $\grainsize$, mass $m_{\rm grain}\equiv (4\pi/3)\,\internaldensity\,\grainsize^{3}$ (in terms of the internal grain density $\internaldensity$), dimensionless absorption+scattering efficiency $\langle Q \rangle_{\rm ext}$, speed of light $c$, and radiation field ${\bf G}_{\rm rad} \equiv {\bf F}_{\rm rad} - \dustvel \cdot (e_{\rm rad}\,\mathbb{I} + \mathbb{P}_{\rm rad})$ in terms of the radiation flux/energy density/pressure density ${\bf F}_{\rm rad}$, $e_{\rm rad}$, $\mathbb{P}_{\rm rad}$. The dust is initialized with the local homogeneous steady-state equilibrium drift and a spatially-uniform dust-to-gas ratio $\initvalupper{\dustden} = \dustgas\,\initvalupper{\gasden}$. For all forces ``from gas on dust'' $a_{\rm gas,\,dust}$ the gas feels an equal-and-opposite force (``back-reaction''). The dust gyro time is given in terms of the grain charge $q_{\rm grain} = Z_{\rm grain}\,e$ as $\tL \equiv m_{\rm grain}\,c / |q_{\rm grain}\,{\bf B}|$, and for the parameter space of our study the drag is given by Epstein drag (as opposed to Coulomb or Stokes drag) with 
\begin{align} \label{eq:ts}
\ts &\equiv \sqrt{\frac{\pi \gamma}{8}}\frac{\internaldensity \,\grainsize}{\gasden\,\cs}\, \bigg( 1+\frac{9\pi\gamma}{128} \frac{|\driftvel|^{2}}{\cs^{2}} \bigg)^{-1/2},
\end{align}

We adopt a standard empirical \citet{mathis:1977.grain.sizes} power-law grain size spectrum with differential number $d N_{\rm d} / d \grainsize \propto \grainsize^{-3.5}$ with a range of a factor of $100$ in grain size ($\grainsize^{\rm max} =100\,\grainsize^{\rm min}$). We assume the grain internal density/composition is independent of grain size, and assume the charge-to-mass ratio scales as $|\graincharge|/\grainmass \propto \grainsize^{-2}$, consistent with grains charged by a range of processes relevant in this regime such as collisions, Coulomb, photo-electric, or electrostatically-limited processes \citep{draine:1987.grain.charging,tielens:2005.book}. 

As in \citet{hopkins:2021.dusty.winds.gmcs.rdis}, we consider two different treatments of the radiation fields. Given the range of column densities we will explore, we are interested in the multiple-scattering regime, or equivalently Rayleigh scattering. In this regime, the radiation should be in the long-wavelength limit (spectrum peaked at wavelengths $\lambda_{\rm rad}\gg \grainsize$), so we expect and assume the spectrally-averaged $\langle Q \rangle_{\rm ext} \propto \grainsize$, and we approximate the radiation with a single band (spectrally-integrated), so effectively treat the grains as introducing a grain size-dependent but otherwise ``grey'' isotropic scattering opacity.  In our first simplified treatment (our ``constant flux'' simulations), we assume the radiation fields obey their homogeneous equilibrium solution, giving ${\bf G}_{\rm rad} \approx {\bf F}_{\rm rad} \approx {\bf F}_{0} = F_{0}\,\hat{z}$. This is a reasonable approximation so long as the radiation is not ``trapped'' in highly-inhomogeneous dust clumps. But we also run a subset of ``full radiation-dust-magnetohydrodynamic'' (RDMHD) simulations where the radiation field is explicitly evolved using to the full M1 radiation-hydrodynamics treatment in {\small GIZMO} \citep{lupi:2017.gizmo.galaxy.form.methods,lupi:2018.h2.sfr.rhd.gizmo.methods,hopkins:2019.grudic.photon.momentum.rad.pressure.coupling,hopkins:radiation.methods,grudic:starforge.methods}, including terms to $\mathcal{O}(v^{2}/c^{2})$: $\partial_{t} e_{\rm rad} + \nabla \cdot {\bf F}_{\rm rad} = -R_{\rm dust}\,\dustvel \cdot {\bf G}_{\rm rad}/c^{2}$, $\partial_{t} {\bf F}_{\rm rad} + c^{2}\,\nabla \cdot \mathbb{P}_{\rm rad} = -R_{\rm dust}\,{\bf G}_{\rm rad}$, where the absorption/scattering coefficients $R_{\rm dust}$ are calculated directly from the explicitly-resolved dust grain populations (consistent exactly with the radiation flux they see in ${\bf a}_{\rm rad}$). 

Our default simulation parameter survey adopts $10^{6}$ gas cells and $4\times10^{6}$ dust super-particles. And unless otherwise specified, our analysis uses the ``full RDMHD'' simulations. Readers interested in details should see \citet{hopkins:2021.dusty.winds.gmcs.rdis}. In that paper, we applied these numerical methods to simulations of radiation-dust-driven outflows in molecular clouds and HII regions. The key differences are (1) we consider a very different parameter space (much higher densities and stronger radiation fields), which lead to qualitatively different instabilities and behaviours, and (2) we specifically model the multiple-scattering regime, while \citet{hopkins:2021.dusty.winds.gmcs.rdis} focused only on the single-scattering limit.

\subsection{Parameter Choices}

Our simulations are then specified by a set of constants (size and charge of the largest grains, dust-to-gas ratio, radiation flux, etc.). To motivate these, we consider a fiducial case of dust around a bright quasar. We expect the most dramatic effects of radiation on dust at the distances closest to the black hole where grains can survive, i.e. just outside the dust sublimation radius $r_{\rm sub} \sim (L_{\rm QSO} / 4\pi\,\sigma_{\rm SB}\,T_{\rm sub}^{4})^{1/2}$ where $T_{\rm sub} \sim 2000\,$K is the dust sublimation temperature and we will consider a typical quasar with $L_{\rm QSO} \sim 10^{46}\,{\rm erg\,s^{-1}}$ (i.e. $M_{\rm B} \sim -24$, a typical $\sim L_{\ast}$ or modestly sub-$L_{\ast}$ QSO at redshifts $z\sim 1-6$; see \citealt{shen:bolometric.qlf.update}), so $r_{\rm sub} \sim 0.3\,$pc and this corresponds to a BH of mass $M_{\rm BH} \sim 10^{8}\,M_{\odot}$ accreting near its Eddington limit.

We then take $\Lscale \sim r_{\rm sub}$, $F_{0} \sim L_{\rm QSO}/(4\pi\,r_{\rm sub}^{2})$, $g \sim G\,M_{\rm BH}/r_{\rm sub}^{2}$, typical $\internaldensity \sim 1.5\,{\rm g\,cm^{-3}}$ and absorption efficiency for the largest grains $\langle Q \rangle_{\rm ext}(\grainsize = \grainsize^{\rm max}) \sim 0.2$ \citep{1984drainelee}, and initial magnetic field strength given by a plasma $\beta_{0} \equiv (\cs / v_{A}[z=0])^{2} = 4\pi\,\rhobase\,(\cs/B_{0})^{2} \sim 1$ with an arbitrary angle $\Bangle^{0} = \pi/4$ (though this is essentially a nuisance parameter here). Observational constraints suggest the dust-to-gas ratios integrated along AGN lines of site range from 0.01-1 times the galactic values \citep{Burtscher2016, Maiolino2001, esparza2021dust}. However, these measurements include regions within the dust sublimation radius and therefore should be interpreted as lower limits. Several studies suggest that the Broad-Line region (BLR) has super-solar dust-to-gas ratios \citep{nenkova2008modeling, kishimoto2009broad, sturm2006hot}. Therefore, given these uncertainties, we assume a standard (galactic) dust-to-gas ratio $\dustgas=0.01$. Further, we consider various values of $\grainsize^{\rm max}$ from $0.01\,{\rm \mu m}$ (smaller grains than typical in the diffuse ISM) through $1\,{\rm \mu m}$ (larger), and also explore variations in the gas density parameterized via the gas column density integrated through the box to infinity, $N_{\rm H,\,gas} \equiv m_{p}^{-1}\,\int \gasden^{0}\,dz = \rhobase\,\Lscale/m_{p} \sim 10^{22}-10^{26}\,{\rm cm^{-2}}$, representative of observed values through different lines-of-sight of angles through the AGN torus.

The one remaining parameter is the dust charge. We have considered both (a) cases where the grains are strongly shielded and the gas neutral/cold, so collisional charging dominates, and (b) cases where some photo-electric (non-ionizing UV) flux can reach the grains. Given the scalings for grain charge in both regimes \citep{tielens:2005.book, draine:1987.grain.charging}, if even a small fraction of the QSO photoelectric flux reaches the grains, they will generally reach the electrostatic photoelectric charging limit such that the equilibrium grain charge $\langle Z_{\rm grain} \rangle \sim 5000\,(\grainsize / {\rm \mu m})$ \citep{tielens:2005.book}. For simplicity, we adopt this by default. However, we note that using the collisional charge expression from \citet{draine:1987.grain.charging}, which results in a significant decrease in $|Z_{\rm grain}|$, has little effect. This is because we find that in the parameter space of interest, the magnetic grain-gas interactions (grain charge effects) are sub-dominant, even with the larger $|Z_{\rm grain}|$. In Appendix \ref{appendix:sims}, we provide a table that lists the specific parameters for each simulation.

%%%%%%%%%%%%%%%%%%%%%%%%%%%%%%%%%%%%%%%%%%%%%%%%%%

\section{Analytic Expectations \& Background}
\citet{hopkins:2017.acoustic.RDI} analyzed the equations of mass and momentum conservation using a linear stability approach to investigate the behaviour of an unstable RDI mode in a dust-gas mixture similar to those simulated in our study. They found that the behaviour of an unstable mode with wave-vector $ \bf k$ is characterized by the dimensionless parameter $\mathbf{k} \cdot \mathbf{w_s} \langle t_s \rangle$, where $\langle {t_s} \rangle = t_s(\langle \rho_g \rangle, \langle \mathbf{w_s} \rangle)$ corresponds to the stopping time at the equilibrium gas density $\langle \rho_g \rangle$ and equilibrium drift velocity $\langle \mathbf{w_s} \rangle$ of the dust particles. This parameter represents the ratio of the dust stopping length to the wavelength of the mode, and defines three regimes of the instabilities, 

\begin{align}
    \label{eq:class}
 \begin{cases}
    \mathbf{k} \cdot \mathbf{w_s}\,   \langle t_s \rangle \lesssim \dustgas \hfill \text{(Low-k, long-wavelength)} \\    
     \dustgas \lesssim \mathbf{k} \cdot \mathbf{w_s} \, \langle t_s \rangle \lesssim (\dustgas)^{-1} \hspace{0.4cm} \text{(Mid-k, intermediate wavelength)} \\
    \mathbf{k} \cdot \mathbf{w_s} \,  \langle t_s\rangle \gtrsim (\dustgas)^{-1}  \hfill \text{(High-k, short-wavelength)}.
\end{cases}
\end{align}
separated by their linear growth rate scaling and mode structure. The different regimes can be further understood by considering the parameter $ \dustgas\,  \mathbf{k}\cdot \mathbf{w_s} \langle t_s \rangle $, which can be interpreted as the ratio of the force exerted by the dust on the gas to the gas pressure forces for a given scale $| \mathbf{k}|$ \citep{moseley:2018.acoustic.rdi.sims}.  The mid-k and high-k regimes exhibit similar behaviour and occur when the gas pressure dominates the dynamics on the scales being considered. Therefore, the resonant mode occurs when the drift velocity aligns with the propagation direction of the gas mode, as given by $\hat{\mathbf{k}}\cdot\mathbf{\driftvel}=\pm c_s$. On the other hand, the low-k regime arises when the bulk force exerted by the dust on the gas becomes stronger than the gas pressure forces, and the dust dominates the flow. Resonant modes in this regime typically align with $\mathbf{\driftvel}$. 

As shown in Equation \ref{eq:class}, the dust-to-gas ratio plays an important role in distinguishing the different RDI regimes. However, for most of our simulations, transitioning into a different regime would require a significant adjustment of $\dustgas$ by several orders of magnitude. Given the specific environmental conditions we aim to model and the likelihood of accurately representing the intended scenario while having such drastic variations in $\dustgas$, we choose to use our fiducial value for $\dustgas$ in all simulations. For a study of the effect of varying $\dustgas$ on the behaviour of the RDIs, we refer readers to \citet{moseley:2018.acoustic.rdi.sims}.

Rewriting the regimes above in terms of wavelength, we can see that $\lambda_{\rm crit} \sim(\internaldensity\,\grainsize) / (\dustgas \, \rhogas) \sim \sizeparam \:\rm \Lscale/ \dustgas$ defines the critical wavelength above which modes are in the low-k regime, where $\sizeparam\equiv (\internaldensity\,\grainsize) / (\rhobase\,\Lscale)$ is the dimensionless grain size parameter which characterizes the coupling strength between the dust and gas. For the parameter set explored here, $\tilde{\alpha} \ll \dustgas$, we find that largest-wavelength interesting modes ($\rm \lambda \sim \Lscale \gg \lambda_{\rm crit}$) always lie in the "long-wavelength" regime. Within the linear theory framework, this mode behaves as a "compressible wave", with similar dust and gas velocity perturbations that are nearly in phase and parallel to the wave-vector $\bf \hat{k}$. This will therefore drive relatively weak dust-gas separation with respect to other regimes previously studied in \citet{hopkins:2021.dusty.winds.gmcs.rdis}. 
The linear growth timescale $t_{\rm grow}$ of the fastest growing modes in this regime scales approximately as:
\begin{align}
\label{growthrate}
     { t_{\rm grow}}(k) \sim \frac{1}{\mathfrak{F}(k)} \sim \Big(\frac{\dustgas \langle \driftvel^2 \rangle k^2}{\langle \ts \rangle} \Big)^{-1/3}, 
\end{align}
where ${\mathfrak{F}(k)}$ is the linear growth rate for a mode with wave-number $k$ \citep{hopkins:2017.acoustic.RDI}. Importantly, as shown therein, the fastest growing mode in the linear long-wavelength regime is the ``pressure-free'' mode, which is weakly dependent on the magnetization and thermal physics of the gas. We discuss this further below.

We define the geometrical optical depth $\tau_{\rm geo}$ instead of the ``observed'' optical depth $\tau_{ \lambda}$ since the latter depends on the observed wavelength (the same integral replacing $\pi \grainsize^2 \rightarrow Q_{\lambda} (\grainsize, \lambda) \, \pi \grainsize^2$), integrated from the base of the box to infinity. Assuming a vertically stratified environment and dust grains with a power-law grain size spectrum, we can express $\tau_{\rm geo}$ strictly in terms of our simulation parameters, 

\begin{align}
    \nonumber \tau_{\rm geo} &\equiv \int_0^\infty \pi \epsilon^2 \, n_{\rm grain} \,dz  \\
   &= C \, \dustgas \, \frac{ \gasden H_{\rm gas}}{\dustden \grainsizemax} =  C \Bigg ( \frac{\dustgas}{\tilde{\alpha}_{\rm m}} \Bigg ),
\end{align}

where $n_{\rm grain}$ is the number density of dust grains, $\sizeparammax$ is the dimensionless maximum grain size parameter ($\sizeparam$ evaluated at $\grainsize=\grainsizemax$), and $C$ is a constant of order 20.  

Another useful parameter is the ``free streaming length'' of the dust (relative to the gas),

\begin{align}
\label{eq:streaming}
\frac{\ell_{\rm stream,\,dust}}{\Lscale} \sim 10^{-4}\,\left( \frac{\grainsize}{\rm \mu m} \right) \, \left( \frac{10^{24}\,{\rm cm^{-2}}}{N_{\rm H,\,gas}} \right) \propto \tau_{\rm geo}^{-1}.
\end{align}

Therefore, for all our simulations, the grains are ``well-coupled'' to the gas in the sense that $\ell_{\rm stream,\,dust} \ll \Lscale$, so we do not expect them to simply ``eject'' from the gas without interacting and sharing momentum.

\subsection{Parameters \& Physics with Weak Effects}

We now discuss physical parameters that we tested, but found to have weak to no effect on the behaviour of the instabilities within this regime including magnetic field strength, magnetic field direction, AGN luminosity, grain charge, and strength of gravity. 

\subsubsection{Charging Physics \& Magnetic Field Strength}
We ran tests varying the magnetic field strength $B_{0}$, or equivalently the plasma $\beta$, and magnetic field orientation $\Bangle$ within the box. Similarly, as the grain charge is unconstrained, we consider different grain charging mechanisms (collisional vs. photoelectric) and found these parameters to have a negligible effect on the long-term behaviour of the instabilities. This is due to two reasons. Firstly, this arises naturally within AGN-like environments where Lorentz forces are weak relative to the drag force, i.e., $\ts / \tL \sim \tilde{\phi}_{\rm m} / \tilde{a}_{\rm d}^{1/2} \ll 1$ where $\chargeparam \equiv 3\,\initvalupper{ \grainchargeZ }[\grainsizemax] \,e / (4\pi\,c\,(\grainsizemax)^{2}\,\rhobase^{1/2})$ is the dimensionless grain charge parameter, and  $\accparam \equiv (3/4)\,(F_{0} \langle Q \rangle_{\rm ext}\, /c )/(\rhobase\,\cs^{2})$ is the dimensionless dust acceleration parameter. Secondly, the dominant modes in our simulations are in the ``long-wavelength regime'', and hence, are only weakly sensitive to magnetic effects as the magnetic pressure and tension provide only second-order corrections to what is to leading order a ``collisionless'' or ``pressure-free'' mode \citep{hopkins:2018.mhd.rdi}. Therefore, we observe that at early stages of the RDIs' development, amplified magnetic fields, or higher grain charge-to-mass ratios merely result in density perturbations propagating at slightly different angles $\sim \Bangle$, but the fluid flow retains its general properties. Further, as the instabilities reach the non-linear stage of their evolution, this propagation angle decreases till the fluid is moving roughly parallel to the vertical acceleration, and we see essentially no effect on the medium. 

\subsubsection{Thermal State of Gas}
We find that the choice of the thermal equation-of-state of the gas $\gamma$, and therefore the speed of sound $\cs$ do not affect our results. As the grains are accelerated to super-sonic velocities, $\cs$ factors out of the relevant equations such as the stopping time and the growth rates of the modes to leading order in the linear theory for these particular long-wavelength modes of interest.

\subsubsection{Gravity}

Further, as shown in Table \ref{table:sims}, for this environment, the strength of gravity is much weaker than the acceleration due to radiation, i.e., $ \tilde g/ \tilde a_d \sim 10^{-3} (\grainsizemax/\micron)$, where $\gravparam \equiv |{\bf g}|\,\Lscale/\cs^{2}$ is the dimensionless gravity parameter and $\accparam \equiv (3/4)\,(F_{0} \langle Q \rangle_{\rm ext}\, /c )/(\rhobase\,\cs^{2})$ is the dimensionless acceleration parameter. Thus, gravity acts merely to ensure that the gas that is left behind the wind ``falls back'', but does not have a noticeable effect on the general behaviour of the RDIs. It is easy to verify that for the conditions and timescales we emulate here, the self-gravity of the gas should also be unimportant.

\subsubsection{AGN Luminosity}

Naively, the AGN luminosity should have an important effect here. However, in the dimensionless units in which we will work, i.e. length in units of $\sim \Lscale \sim r_{\rm sub}$, time in units of the ``acceleration time'' defined below, the absolute value of the AGN luminosity factors out completely. Nonetheless, while the AGN luminosity does not affect the qualitative behaviour of the RDIs (in the appropriate units), it effectively defines the characteristic time and spatial scales of the problem. For example, the AGN luminosity normalizes the sublimation radius, i.e. $r_{\rm sub} \sim 0.3 \,{\rm pc}\,L_{46}^{1/2}$. This means if we define the flux at the base of our box as the flux at $r_{\rm sub}$ (as we do), the AGN luminosity factors out (the flux at $r_{\rm sub}$ is, by definition, fixed \citep{ivezic1997self}), and we find that the vertical acceleration of the column, $ a_{\rm eff} \equiv \dustgas a_{\rm dust} - g \sim a_{\rm eff} \equiv \dustgas a_{\rm dust}$, where $a_{\rm dust}$ is the acceleration experienced by the dust, has the following scaling, 

\begin{align}
     a_{\rm eff} \sim 0.3 \, \text{cm}\, \text{s}^{-2} \Bigg ( \frac{1 \micron}{\grainsizemax} \Bigg ),
\end{align}
which is independent of the AGN luminosity, and only depends on the maximum size of the grains. 

It is worth noting that our choice of normalization is not arbitrary.  In the context of dust-driven winds, our focus is on regions where dust is present, i.e., beyond the sublimation radius. When the radius is much smaller than the sublimation radius ($r \ll r_{\rm sub}$), the dust is expected to be sublimated, and the dominant mechanism for driving the wind would be line-driving rather than dust absorption \citep{proga2000dynamics}. Conversely, when the radius is much larger than the sublimation radius ($r \gg r_{\rm sub}$), the radiation flux decreases according to the inverse square law. In our simulations, we observe that the wind originates from the base of the column where the radiation flux is strongest, which aligns with our expectations. The sublimation radius can be derived analytically by assuming thermal equilibrium, allowing allows us to express the sublimation radius as $r_{\rm sub} \sim (L_{\rm QSO} / 4\pi\,\sigma_{\rm SB}\,T_{\rm sub}^{4})^{1/2}$. Therefore, since the location of the dusty torus is proportional to $\sqrt{L_{\rm QSO}}$, the flux at the inner edge of the torus is independent of luminosity. This size-luminosity relation has been supported by observational studies \citep{tristram2009parsec,kishimoto2011mapping, suganuma2006reverberation}. However, it is important to note that the theoretical relation strongly depends on the sublimation temperature, which in turn depends on grain composition which is uncertain. In our simulations, we assume a silicate grain composition corresponding to a sublimation temperature of 1500 K. Nevertheless, different grain compositions within the torus can result in sublimation temperatures ranging from  $\sim 1300$ K to $2000$ K. This variation influences the flux and acceleration timescales of the winds, resulting in a fractional variation of 0.6 for the sublimation radius, where smaller (larger) radii would correspond to shorter (longer) timescales for wind launching.

However, the argument above assumes that the flux is stronger than the gravitational pull of the central source, allowing the initiation of a wind. Therefore, the luminosity does not affect the behaviour of the wind insofar as this condition is met. 

The luminosity does however, normalize the bulk acceleration timescale which depends on both $\Lscale \sim r_{\rm sub}$ and $a_{\rm eff}$, as
\begin{align}
\label{eq:tacc}
    \nonumber t_{\rm acc} &\equiv \sqrt{\frac{20 H_{\rm gas}}{a_{\rm eff}}}\\
    &\sim 245 \,\text{yrs} \, L_{46}^{1/4}\,\Bigg(\frac{\grainsizemax}{ \micron}\Bigg)^{1/2} \,\Bigg(\frac{0.01}{\dustgas}\Bigg)^{1/2},
\end{align}
corresponding to the time when a perfectly coupled dust + gas fluid would have reached a height $z\sim 10\,\Lscale$. As we normalize our parameters to the sublimation radius $r_{\rm sub}$ and the bulk acceleration timescale $t_{\rm acc}$, our findings are independent of the AGN luminosity. However, if the dust were held at a fixed radius while varying the luminosity, the flux at the sublimation radius would change, which could alter the dynamics of the fluid and thus, affect the behaviour of the RDIs.

\subsection{Parameters with Strong Effects: The Geometric Optical Depth}
Our results are sensitive to the choice of grain size and column density, as they determine the critical wavelength and thus the dominant mode of the instability.  Specifically, from Equation \ref{eq:class}, we can see the ratio of the largest scale mode with $\lambda \sim \Lscale$ to critical wavelength can be expressed as
\begin{align}
\label{eq:length}
    \nonumber \frac{\Lscale}{\lambda_{\rm crit}} &\sim \frac{\Lscale}{ \langle \driftvel \rangle \, \ts /\dustgas} \sim \frac{\dustgas \Lscale}{\internaldensity}\Bigg (\frac{\rhogas}{\grainsize}\Bigg )\sim \frac{\dustgas}{\tilde \alpha_{\rm m}} = \frac{\tau_{\rm geo}}{C}\\
     & \sim 300 \, \left(\frac{\dustgas}{0.01}\right) \left (\frac{\rm N_H }{ 10^{24} \, \rm cm^{-2}}\right) \left(\frac{1 \, \micron }{\grainsizemax} \right), 
\end{align}

where $C\sim 20$ is a constant defined earlier. 

Again, as $\Lscale/\lambda_{\rm crit}\gg 1$ for the typical values of ($\rhogas/ \grainsizemax$), the dominant modes are always in the long-wavelength regime. Additionally, we note the regime of the instabilities strictly depends on the geometrical optical depth, where an environment with $\tau_{\rm geo} \gtrsim 20$ would be sufficient to satisfy the criteria for the ``long-wavelength RDI '' regime.

Further, we can compare the instability growth time to the wind's acceleration time. As $a_{\rm eff} \gg \cs / \ts^{0}$, where $\ts^{0}$ is the stopping time at $t=0$, we assume that the dust is drifting super-sonically and use the expression for the equilibrium drift velocity in the supersonic limit derived in \citet{hopkins:2017.acoustic.RDI} (i.e., $\langle \driftvel \rangle \sim \sqrt{a_{\rm dust} \ts^{0} \cs}$) with direction $\driftvelhat$ to obtain

\begin{align}
\label{eq:time}
    \nonumber \frac{t_{\rm acc}}{ t_{\rm grow}} &= \left ( \frac{20 \Lscale}{a_{\rm dust}}\right )^{1/2}\left(\frac{(\mathbf{k\cdot \driftvel})^4}{\dustgas \langle \ts \rangle ^2} \right)^{1/6}, \\ &\sim 4.7 \, (\Lscale\,{\bf k \cdot \bf \driftvelhat)}^{2/3}\,\Bigg (\frac{\tilde \alpha}{\dustgas}\Bigg)^{1/6} \propto \tau_{\rm geo}^{-1/6}.
\end{align}

Note that $\Lscale\,{\bf k} \cdot {\bf \driftvelhat} \sim 1$ and that $\dustgas/\tilde{\alpha}_{\rm m} \sim \tau_{\rm geo}/C$. Hence, the characteristic timescales and length scales only depend on $\tau_{\rm geo}$ or the ratio $\dustgas N_{\rm H}/\grainsizemax$, yielding similar behaviours for similar ratios. As $t_{\rm acc}/ t_{\rm grow} \propto \tau_{\rm geo}^{-1/6}$, lower $\tau_{\rm geo}$ (lower column density and larger grains) imply shorter growth times, i.e. more e-folding times for the clumping to amplify. This would result in filaments with stronger clumping and higher variability. However, we note that this trend is weak $\sim \tau_{\rm geo}^{1/6}$, so we observe similar levels of clumping/variability across the parameter space we explore. 

From the relations obtained in Equations \ref{eq:length} and \ref{eq:time}, it is evident that $\dustgas$ plays a crucial role in shaping the spatial and temporal behaviour of the RDIs. In our simulations, we have employed a fixed value of $\dustgas = 0.01$. However, it is important to recognize that this parameter will vary depending on the AGN environment and metallicity $Z$. The connection between $\dustgas$ and $Z$ is derived based on the assumption that dust formation and destruction timescales exhibit similar dependencies on time \citep{dwek1998evolution}. To first-order, this leads to a constant dust-to-metal mass ratio and a dust-to-gas ratio that scales with metallicity as $\dustgas \propto Z$, which is supported by observational studies \cite[e.g.,][]{draine2007dust, james2002scuba, bendo2010herschel, magrini2011herschel}. For $\dustgas \gg 0.01$, we anticipate minimal deviations in RDI behaviour, as the RDIs would still reside within the long wavelength regime. Although the ratio $t_{\rm acc}/t_{\rm grow}$ would would be reduced according to $t_{\rm acc}/t_{\rm grow} \propto (\dustgas)^{-1/6}$, the impact is not substantial. However, increasing $\dustgas$ would result in a higher dust opacity, thereby requiring a lower UV luminosity to initiate outflows. In addition, these outflows would have shorter acceleration times ($t_{\rm acc} \propto (\dustgas)^{-1/2}$). In environments where $\dustgas \ll 0.01$, a shift in RDI behaviour may occur. Specifically, in low density columns ($\rm N_H \leq 10^{22} \, \text{cm}^{-2}$) with maximum grain sizes $\grainsizemax \geq 1 \, \mu\text{m}$, the RDIs could transition to the mid-wavelength regime due to the linear dependence of $\Lscale/\lambda_{\rm crit}$ on $\dustgas$.

However, in order to induce significant changes in RDI behaviour driven by variations in metallicity or the dust-to-gas ratio, $\dustgas$ would need to undergo a shift of at least one order of magnitude. Observations suggest that the majority of AGN environments exhibit solar-to-supersolar  metallicities \citep{hamann2002metallicities, storchi1998chemical}. Low-metallicity AGN sources have been observed, however, they only display marginal deviations below solar metallicity \citep{polimera2022resolve, ubler2023massive, groves2006emission}.

\section{Results}
\label{sec:results}
\subsection{General Profile of the Outflow and Large scale Morphology}
\begin{center}
\begin{figure*}
    \centering
    \includegraphics[height=0.94\textheight]{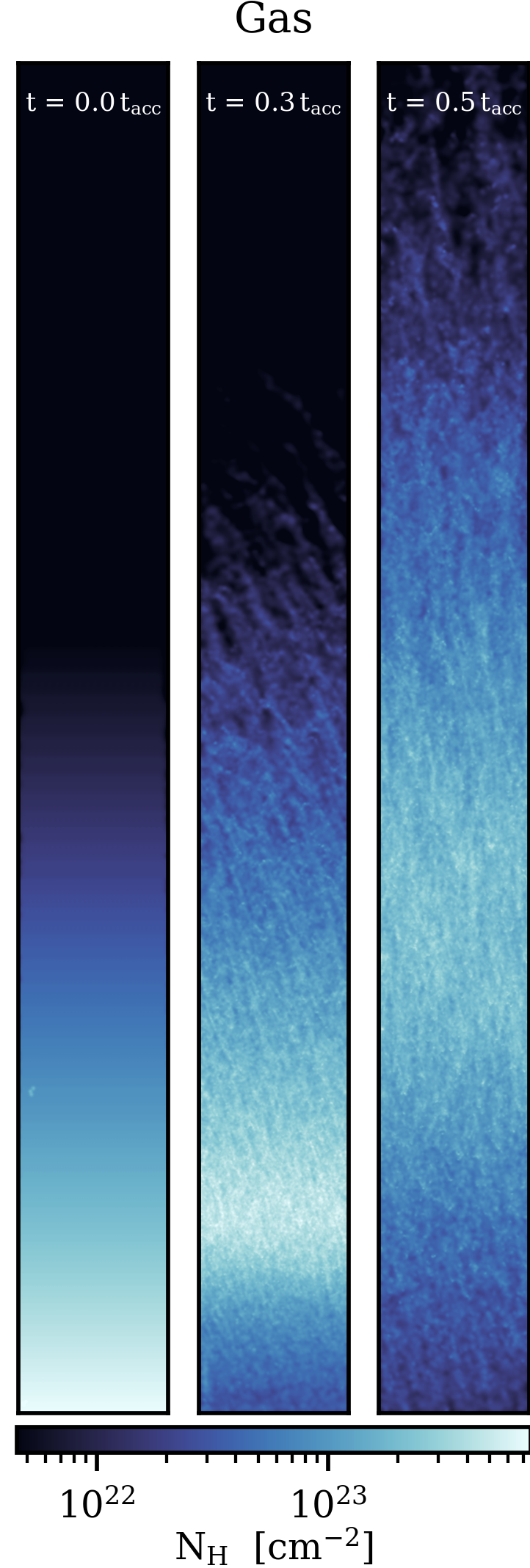}\hfil
    \includegraphics[height=0.94\textheight]{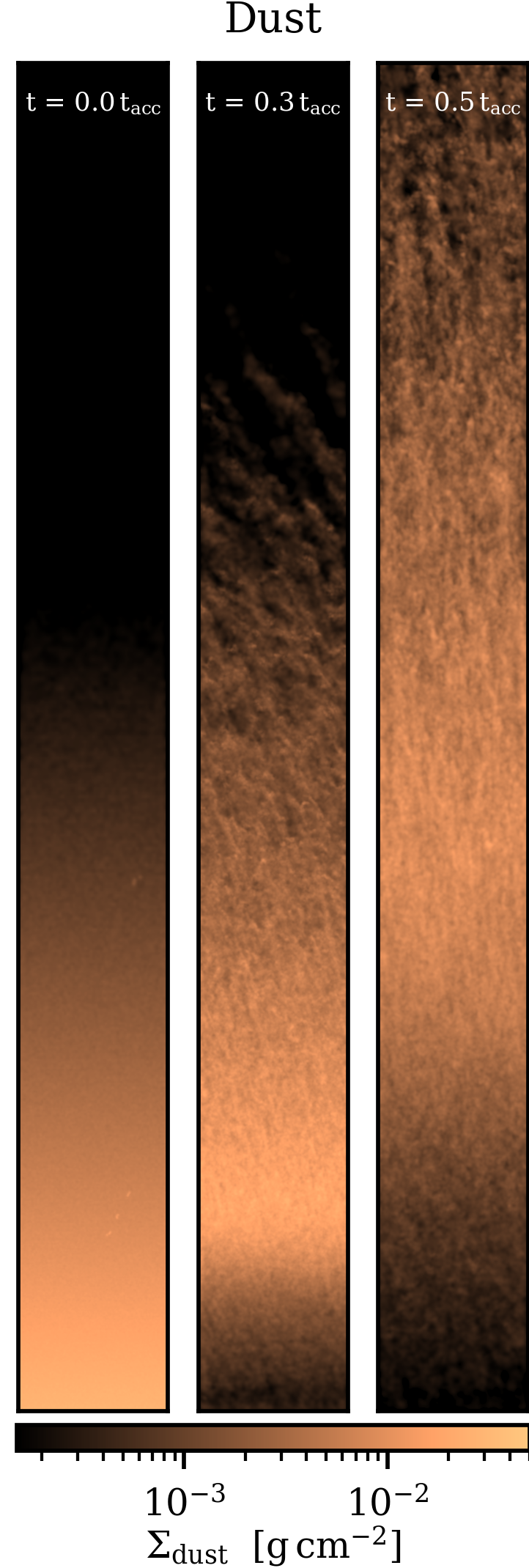}
    \vspace{-0.15cm}
    \caption[width=0.45\textwidth]{The evolution of the gas (left), and dust (right) column density for a simulation box with $\rm N_H \sim 10^{24}cm^{-2}$ and $\grainsizemax \sim 1 \micron$ in the $xz$ plane within $z \sim 0-9 \, \Lscale$ at $t\sim (0.0, 0.3, 0.5) \, t_{\rm acc}$, where $t_{\rm acc}$ corresponds to the  acceleration timescale defined as $t_{\rm acc} \equiv (20\,H_{\rm gas}/\langle a_{\rm eff} \rangle)^{1/2}$ with $\langle a_{\rm eff} \rangle \equiv
    \langle \dustgas a_{\rm dust,\,rad}  \rangle - g$, when a perfectly coupled fluid would have reached a height $z\sim 10\,\Lscale$. All simulations within our set show winds that were successfully launched with high degrees of clumping on small spatial scales and vertical filaments on large scales. The RDIs develop within a fraction of wind acceleration time ($\rm t_{grow} \sim 10^{-1} \, t_{acc}$) with similar structures for the gas and dust. The filaments that form are initially inclined with respect to the $\hat{z}$ direction and align along the $\hat{z}$-axis at later times $(\rm t \sim 0.5 \, t_{acc})$. }
    \label{fig:fullbox_gas_t}
\end{figure*}
\end{center}
To understand how the RDIs affect the dynamics of the dusty torus, we first consider the resulting morphology within a relatively small patch within the torus. However, as we are not modelling the entire region around the AGN, we cannot draw definitive conclusions about how the RDIs affect the overall morphology of the AGN torus or its geometry. The results we present in Figure \ref{fig:fullbox_gas_t} show the temporal evolution of the gas (left) and the dust (right) column densities for a run with $\rm N_{H}\sim 10^{24} cm^{-2}$ and $\grainsize \sim 1 \micron$ in the $xz$ plane within $z \sim 0-9 \, \Lscale$ at $t\sim (0, 0.3, 0.5) \, t_{\rm \text{acc}}$. These plots illustrate the successful launch of a radiation-driven wind with strong gas-dust coupling and the formation of elongated filaments on large scales. At $t \sim 0$, the fluid is vertically stratified as per our initial conditions. The RDIs have growth times that are short relative to the flow time, with the largest scale modes growing at a fraction ($\sim 10^{-1}$) of wind acceleration time. While the instabilities are within the linear regime, the gas and negatively charged dust develop density perturbations in the form of sinusoidal waves at an inclination angle $\sim - \Bangle^0$ from the vertical axis. As the instabilities evolve non-linearly, the inclined filaments begin aligning with the vertical axis forming elongated structures that continue to accelerate upwards. 

% This inclination is driven by a super-position of the radiative pressure and Lorentz forces acting on negatively charged dust grains. At later times as the velocity of the grains is reduced due to drag forces, and thus the ratio of the Lorentz force to radiative flux force, $F_L/F_r$, is reduced, this drives the inclined filaments to begin aligning with the vertical axis forming elongated structures. 

% As discussed in \citet{hopkins:2017.acoustic.RDI}, the resonant modes satisfy the condition, $w_s = \plusminus 1/cos(\theta)$. 

The centre of the wind, which we define as the region containing a dominant fraction of the dust ($60\%$ by mass of the dust within the central region with $20\%$ below and above the region), reaches a height similar to that expected for a perfectly coupled homogeneous fluid without any RDIs present. However, we find that only $\sim 50 \%$ of the gas remains within such heights, with roughly $40 \%$ of the gas ``lagging'' behind the wind. This suggests that while there exists strong micro-scale coupling between the dust and the gas through drag forces, the overall fluid is not perfectly coupled resulting in a significant fraction of the dust ``leading'' in front of the gas.

\begin{center}
\begin{figure}
    \centering
    \includegraphics[width=0.45\textwidth]{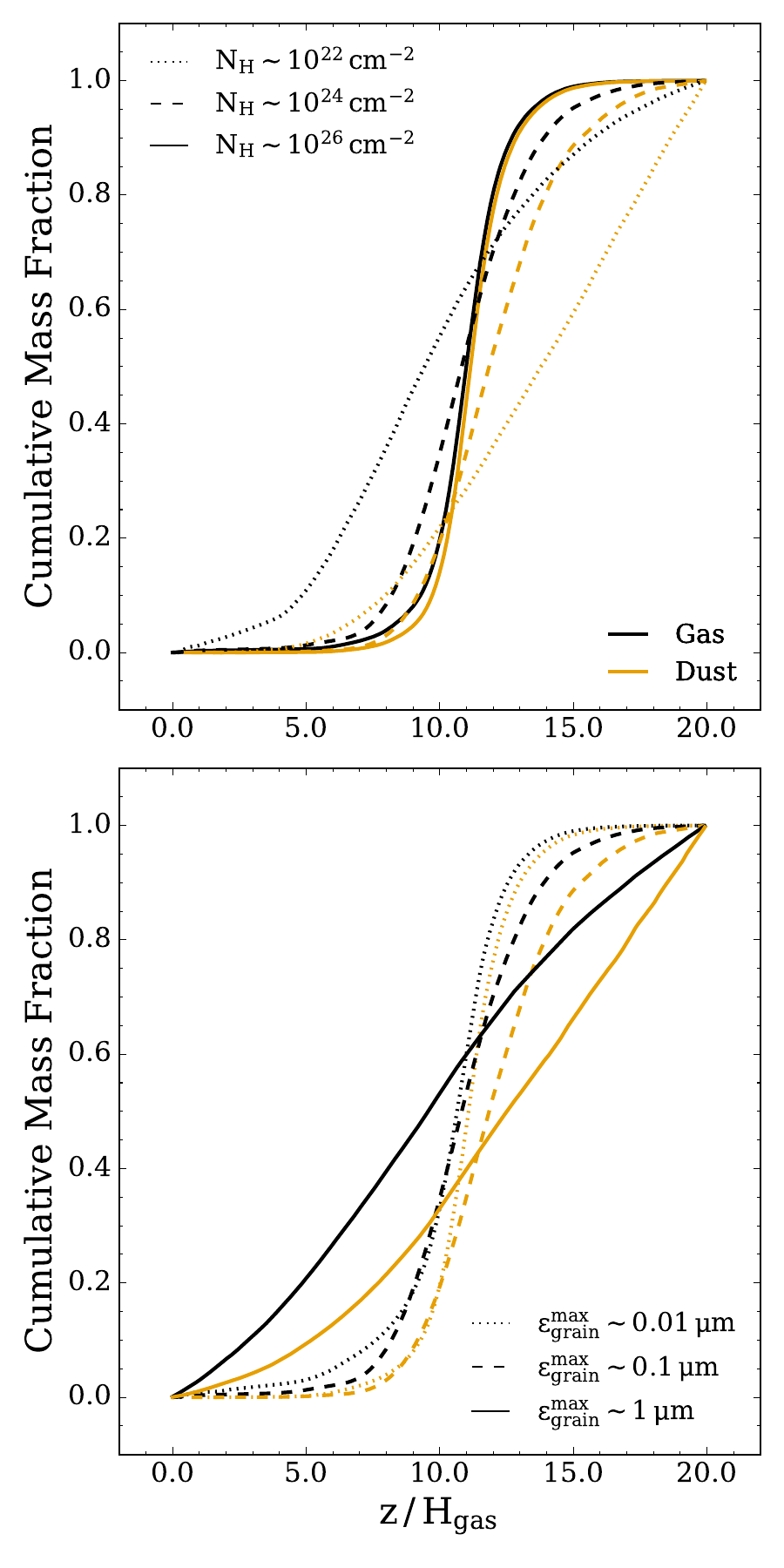}
    \caption{Cumulative mass fraction (CMF) profile of the gas (black) and the dust (yellow) for different column densities and maximum grain sizes.  The top panel displays the results for a fixed maximum grain size of $\grainsizemax \sim 0.1 \micron$ and column densities of $\rm N_H \sim 10^{22}\text{cm}^{-2}$ (dotted line), $\rm N_H \sim 10^{24}\text{cm}^{-2}$ (dashed line), and $\rm N_H \sim 10^{26}\text{cm}^{-2}$ (solid line) at $t \sim t_{\rm acc}$. The bottom panel shows the profiles for a fixed column density of $\rm N_H \sim 10^{24} \text{cm}^{-2}$ and different maximum grain sizes: $\grainsizemax \sim 0.01 \micron$ (dotted line), $\grainsizemax \sim 0.1 \micron$ (dashed line), and $\grainsizemax \sim 1 \micron$ (solid line). At high column densities and small grain sizes, the gas and dust show similar profiles, but the fluid is not perfectly coupled, with the gas "lagging" behind the dust. This decoupling becomes more pronounced at lower column densities and larger grain sizes as $ \ell_{\rm stream,\,dust} \propto \epsilon_{\rm grain} \, \rm N_{\rm H}^{-1}$.}
    \label{fig:massfrac}
\end{figure}
\end{center}

Additionally, we find that the cumulative mass fraction (CMF) profile of the outflow strongly depends on the parameters we explore within our simulation set. As depicted in Figure \ref{fig:massfrac}, we present the CMF profile of the gas and dust at $t\sim t_{\rm acc}$ for different simulations. To demonstrate the effects of varying column densities, the top panel shows the results for simulations with maximum grain size $\grainsizemax \sim 0.1 \micron$, and average column density $\rm N_H \sim 10^{22}\text{cm}^{-2}$, $\rm N_H \sim 10^{24}\text{cm}^{-2}$ and $\rm N_H \sim 10^{26}\text{cm}^{-2}$. Meanwhile, to show the grain size dependence, the bottom panel displays the CMF profile for simulations with an average column density of $\rm N_H \sim 10^{24}\text{cm}^{-2}$ and maximum grain sizes of $\grainsizemax \sim 0.01 \micron$, $\grainsizemax \sim 0.1 \micron$, and $\grainsizemax \sim 1 \micron$. Although the gas and dust have similar CMF profiles, the fluid is not perfectly coupled, with the gas ``lagging'' behind the dust. This lagging effect increases with increasing grain size and decreasing density as predicted in Equation \ref{eq:streaming}. In our runs with higher column densities ($\rm N_H \sim 10^{25} 
- 10^{26} \rm cm^{-2}$), the two plots roughly overlap as the fluid becomes closer to a perfectly coupled fluid on large scales. To measure the impact of imperfect coupling between gas and dust, we analyze the cumulative mass fractions of the gas compared to the dust at $t = t_{\rm acc}$. By comparing the height range that encompasses 25-75\% of the dust to the corresponding gas mass within that range, we can quantify this effect. Our findings show that, on average, the dusty gas can successfully eject around 70-90\% of the gas present. This implies that the torus is not a static or constant structure, but rather subject to substantial variations over time. If a high luminosity state persists for a sufficient duration to drive a wind, it is anticipated that the torus would disappear. This aligns with the receding torus framework as proposed in \cite{lawrence1991relative, simpson2005luminosity, hoenig2007active}.

\subsection{Effects of Full RDMHD}
\label{subsec:radhydro}

In Figures \ref{fig:fullbox_gas_t} \& \ref{fig:rhd_nH}, we compare the morphology of the simulations for our full RDMHD runs\footnote{In these simulations, we can optionally employ a reduced speed of light (RSOL) \citep[see][]{hopkins:2021.dusty.winds.gmcs.rdis}, $\tilde{c}<c$. In tests, we find identical results for $\tilde{c} \sim (0.1 - 1) c$ at $\rm N_H \lesssim 10^{25} \, cm^{-2}$, so we use $\tilde{c} = 0.1 c$ here so we can run at our higher fiducial resolution. For $\rm N_H \gtrsim 10^{25} \, cm^{-2}$, however, finite speed of light effects are important so we use $\tilde{c} = c$ (no RSOL). This imposes a large CPU cost (shorter timesteps), so the full RDMHD simulations of $\rm N_H \gtrsim 10^{25} \, cm^{-2}$ use 10x fewer resolution elements.} (\fref{fig:rhd_nH}) versus the approximate ``homogeneous flux'' ($F_0 =$ constant) simulations (\fref{fig:fullbox_gas_t}). Our RDMHD simulations employ a grey band approach with a photon injection rate of $\sim L/c$ where the optical depth ($\tau_{\rm IR}$) is set to crudely represent the IR opacity of the column. Therefore, the values we present for $\tau_{\rm IR}$ should serve as rough estimates rather than precise values as we do not account for effects like the wavelength dependence of the opacity or photon degradation. From left to right in Figure \ref{fig:rhd_nH}, the simulations correspond to columns with $\rm N_H\sim 10^{22} \, cm^{-2}, 10^{24} \, cm^{-2}, 10^{26} \, cm^{-2}$ respectively and $\grainsizemax \sim 1 \micron$ at $\rm t\sim 0.5 \, t_{\rm acc}$. We discuss the different regimes shown in this figure in the subsections below. 

\subsubsection{Intermediate Optical Depths \& The ``Acceleration limited'' Regime}
\begin{figure*}
    \centering
    \includegraphics[width=0.45\textwidth]{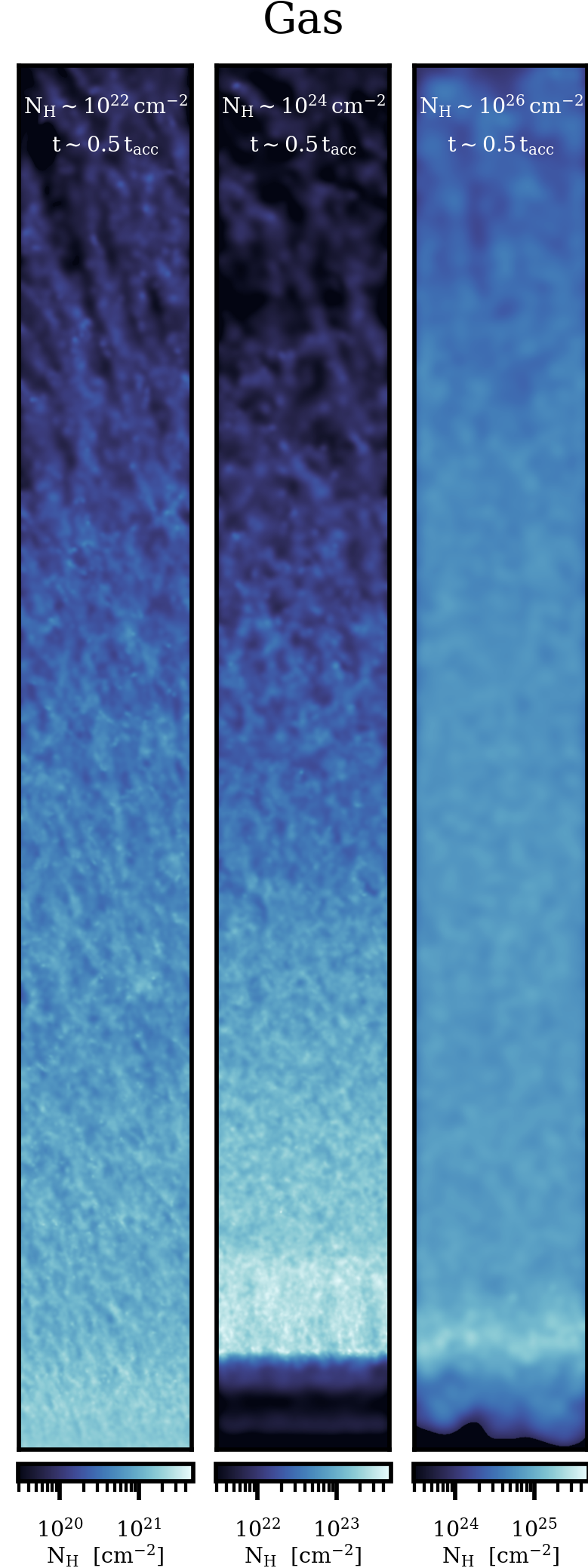}
    \includegraphics[width=0.45\textwidth]{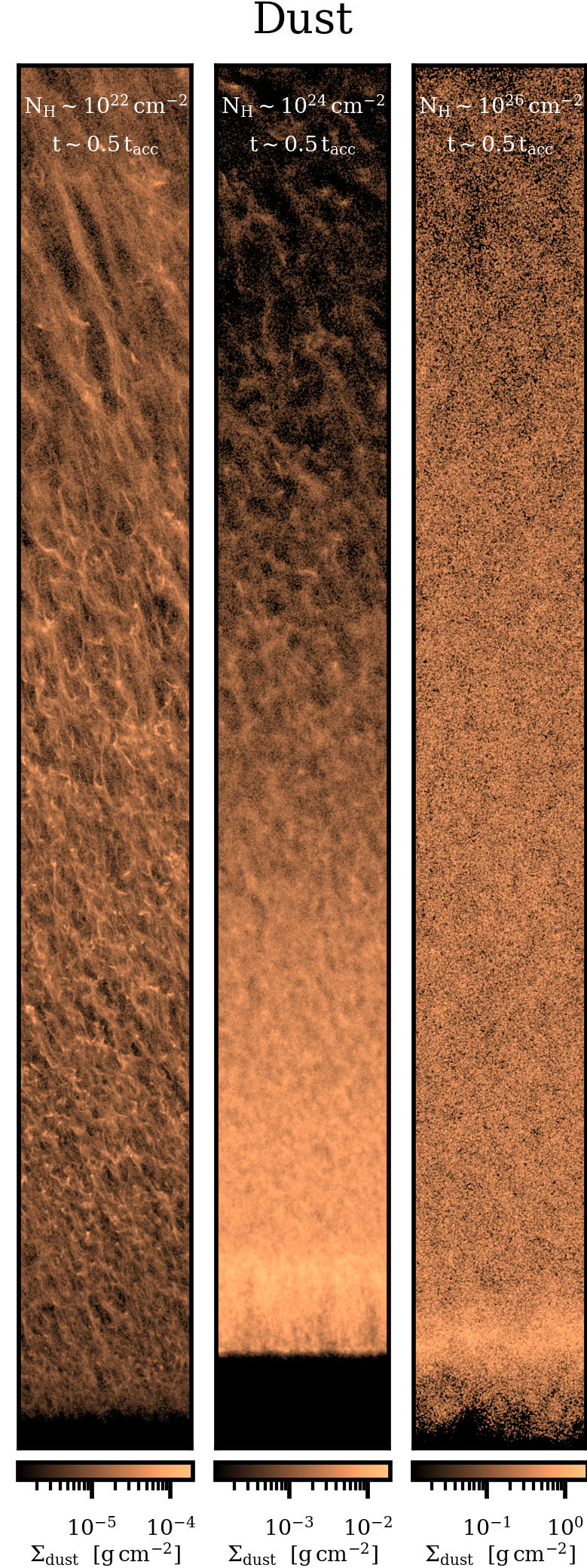}
    \caption[width=0.5\textwidth]{The gas (left) and dust 
    (right) column densities for full RDMHD runs projected onto the $xz$ plane within $z \sim 1-9 \, \Lscale$ at $t\sim 0.5 \, t_{\rm acc}$. From left to right, the simulations correspond to runs with $\grainsizemax \sim 1 \micron$, and  $\tilde c \sim (0.1, 0.1, 1) \, c$, $\rm N_H \sim 10^{22}cm^{-2}, \, 10^{24}cm^{-2}, 10^{26}cm^{-2}$ corresponding to $\tau_{\rm geo} \sim 20, 2000, 2\times 10^5$ ($\tau_{\rm IR} \sim 0.2, 20, 2000$) respectively. Note that the right-most plot shows less small scale structure due to a factor of 10 reduction in resolution (owing to the cost of using $\tilde{c}=c $). For the $\rm N_H \sim 10^{22}cm^{-2} \, \text{and} \,  10^{24}cm^{-2}$, the optical depth is sufficiently low such that full treatment of RDMHD shows similar structure formation on small and large scales to the runs without explicit radiative transfer. For the $\rm N_H \sim 10^{26}cm^{-2}$ column, the high optical depth results in photon diffusion time that are longer than the wind acceleration time expected from a constant flux assumption resulting in a slower outflow.}
    \label{fig:rhd_nH}
\end{figure*}

For this regime, we consider the left and middle panels in Figure \ref{fig:rhd_nH} with $\rm N_H\sim 10^{22}\, cm^{-2}$ and  $\rm N_H\sim 10^{24} \, cm^{-2}$, which correspond to $\tau_{\rm geo} \sim 20$ ($\tau_{\rm IR} \sim 0.2$) and $\tau_{\rm geo} \sim 2000$ ($\tau_{\rm IR} \sim 20$) respectively. With reference to Figure \ref{fig:fullbox_gas_t}, we can see that to first order, the large-scale morphology of the RDIs does not show any significant changes when the simulations are run with our full radiative transfer treatment versus simply assuming a homogeneous radiation field. We do note the formation of a thin high density \textquote{slab} at the base of the box in the middle panel of Figure \ref{fig:rhd_nH}. This ''slab'' acts as an opaque wall that gets lifted by the incident photons, and effectively translates the wind upwards without significant distortions to its morphology. Nonetheless, this \textquote{slab} does not significantly affect the integrated surface density along the line-of-sight or any of the macro-scale properties of the column above it, such as the CMF or clumping factor profile. Therefore, we conclude that using the homogeneous radiation approximation is sufficient within this regime. We emphasize, as shown in the following section, that the key factor is the radiation diffusion time compared to the wind launch and instability growth timescales. When the radiation diffusion time is fast compared to these timescales, the radiation field is smooth, and the homogeneous radiation approximation is valid.

\subsubsection{Extremely Large Optical Depths: The Radiation-Propagation Limited Regime}
\label{tau_thick}
For this regime, we consider the panel on the right in Figure \ref{fig:rhd_nH} with $\rm N_H\sim 10^{26}\,  cm^{-2}$, which corresponds to $\tau_{\rm geo} \sim 2\times 10^{5}$ ($\tau_{\rm IR} \sim 2000$). We point out that the plot displays less small scale structure than the panels on the left due to the reduced resolution of the simulation (as noted above, this owes to using no ``reduced speed of light'' here, which imposes a steep computational cost penalty). For this case, when accounting for full radiative transfer, the fluid is found to be accelerated to a lower height than expected. This result can be attributed to the breakdown of the assumption of an infinitesimally small photon diffusion timescale (constant flux field). As the photons travel through the fluid, they ``lag'' behind the wind due to propagation effects, leading to a decrease in the radiative acceleration and consequently, the fluid being accelerated to a lower height than expected. To determine when this occurs, we consider the ratio of the photon diffusion time, $\rm t_{diff}$, to the dust acceleration time. For simplicity, we ignore the effects of gravity and assume a homogeneous dust-gas distribution. Therefore, the ratio of the time needed for a photon to diffuse through a distance $H_{\gamma}$ (the ``width'' of the gas ``shell'') to the time required to accelerate the same ``shell'' to a height of $10 \, \rm H_{gas}$ has the following scaling,

\begin{align}
\label{eq:ratio}
    \rm \frac{t_{diff}}{t_{acc}} &= %\frac{\ell^2_{\rm stream} \kappa \rhogas}{c}=
    \frac{H_{\gamma}^{2} \, \dustgas \, \rhogas \,  \kappa \,  a_{\rm eff}^{1/2}}{\sqrt{20 \Lscale}c} \\
    \nonumber &=\frac{3}{8\sqrt{5}}
    \, \frac{c_s}{c}\, \langle Q_{\rm ext}\rangle \, \accparam^{1/2}\, \Bigg( \frac{\dustgas}{\tilde \alpha_m} \Bigg)^{3/2} \left( \frac{\rm H_{\gamma}}{\Lscale} \right)^2\\
    \nonumber &\sim 5 \times 10^{4} \Bigg ( \frac{c_s/c}{10^{-5}}\Bigg ) \Bigg ( \frac{\langle Q_{\rm ext}\rangle}{1} \Bigg ) \Bigg (\frac{ \accparam}{5\times 10^7} \Bigg )^{1/2} \Bigg ( \frac{\tau_{\rm geo}}{2 \times 10^4} \Bigg )^{3/2}\left( \frac{\rm H_{\gamma}}{\Lscale} \right)^2
\end{align}
where $\kappa$ is the dust opacity, and we assume that $c_s/c \sim 10^{-5}$ ($\rm T_{gas} \sim 1000\, K$), matching the assumptions used in our simulations. For simplicity, we assume that the grains all have the median grain size ($\grainsize \sim 0.1 \grainsizemax$) and not a grain size spectrum. It is important to note that the expression above is sensitive to the value of $\tau_{\rm geo}$. When comparing our lowest optical depth simulation ($\rm N_H \sim 10^{22} \, cm^{-2}$) to our highest ($\rm N_H \sim 10^{26} \, cm^{-2}$), there is an increase of a factor of $10^{4}$ in $\tau_{\rm geo}$, which in turn results in a factor of $10^{6}$ in the ratio of the two timescales considered above. Therefore, in the higher optical depth case, the radiation can no longer propagate fast enough to reach the material at the top of the box to maintain a constant flux. Consequently, material at the ``top'' of the box in the ICs can fall down before radiation reaches it and the outflow propagation speed is limited not just by naive total acceleration but also photon transport time, resulting in a wind with a slower outflow velocity. However, despite the morphological change on large scales, this effect mostly acts to reduce the vertical translation of material in the column at a given time and has minimal effect on the internal properties of the outflow.

\subsection{Do Winds Launch?}
\label{subsec:launch}
\begin{center}
\begin{figure}
    \centering
    \includegraphics[width=0.45\textwidth]{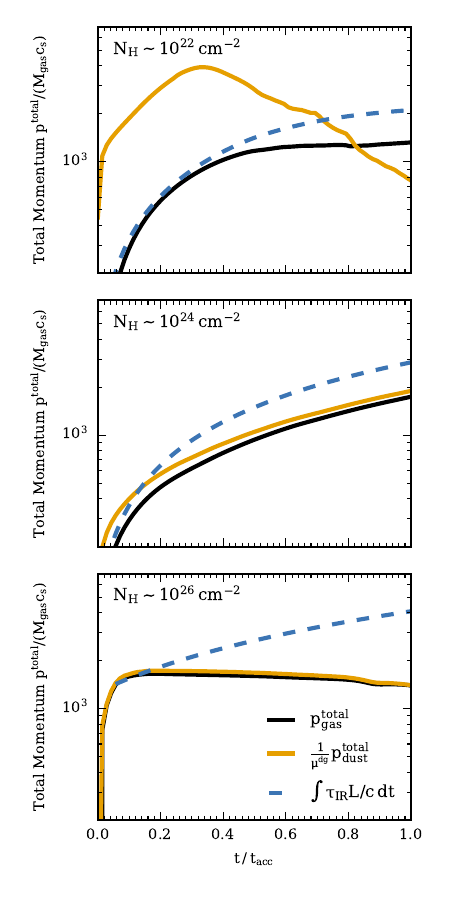}
    \vspace*{-5mm}
    \caption{Gas (black) and dust (yellow) total momentum normalized to the product of the total gas mass within the box at $t=0$ and the speed of sound compared to expected momentum in the wind (blue dotted line) in the homogeneous perfect-coupling grey opacity limit, $\rm p^{total}\sim \int \, \tau_{\rm IR} L/c \, dt$ for 3 RDMHD runs. Note that we multiply the dust momentum by a factor of $1 / \dustgas = 100$ for plotting purposes. From top to bottom, the total gas column density $N_\text{\rm gas}$ corresponds to $10^{22}\text{cm}^{-2}$, $10^{24}\text{cm}^{-2}$, and $10^{26}\text{cm}^{-2}$ respectively, and maximum grain size of $1 \micron$. The plots show factors of $1-3$ momentum ``leakage'', with higher leakage for denser columns. The top panel shows a turnover in the dust momentum as most energetic dust particles escape from the box.}
    \label{fig:tau}
\end{figure}
\end{center}

As shown in Figures \ref{fig:fullbox_gas_t} and \ref{fig:massfrac}, our plots indicate that the accelerated dust imparts sufficient momentum onto the gas to successfully launch a wind across our entire parameter survey.  As photons propagate through the box, they could in principle escape through low density \textquote{channels}, and thus, impart lower amounts of their momentum onto the dust resulting in $\rm p^{total} < p^{MS} \equiv \int \tau_{\rm IR} \rm L/c  \, dt$, where $\rm p^{total}$ and $\rm p^{MS}$ denote the total momentum carried by the fluid and the expected momentum for the multiple scattering regime respectively. We show the gas and dust components of the total momentum (note that we multiply the dust momentum by a factor of $1/\dustgas = 100$ for ease of comparison) in the wind relative to the predicted momentum $\rm p^{MS}$ in Figure \ref{fig:tau}. The plots show that prior to the growth time for the instabilities, $\rm t\lesssim \, 0.1 \, t_{acc}$, the radiation is well coupled to the fluid. However, as the instabilities grow, the line for the expected momentum begins to separate from the imparted momentum as low density \textquote{channels} develop. When the total momentum of the fluid in the simulation is lower than the expected value, we define this as momentum ``leakage''. This situation indicates a lack of efficient momentum transfer between the injected radiation and the dusty fluid. At $\rm t\sim t_{acc}$, the plots show factors $1-3$ of momentum ``leakage'' from the box which increases with increasing column density. We attribute this effect to slower photon diffusion at higher column densities which results in an overall reduced incident flux on the dust particles. But we still always see an order-unity fraction of the radiation momentum $ \rm p^{MS}$ actually couples, and thus is always sufficient to launch a wind under AGN-like conditions as simulated here. 

We compare our simulations to the simulations conducted by \citet{venanzi2020role} and \citet{arakawa2022radiation} modelling AGN winds driven by radiation pressure on dust. A key distinction in our approach is that we explicitly account for dust dynamics, which was not taken into consideration in the previous simulations. Consistent with the findings of \citet{arakawa2022radiation}, we observe that the acceleration of the gas column remains unaffected by column density in the multiple scattering regime ($\rm N_H \sim 10^{22}-10^{24} \, \text{cm}^{-2}$), as indicated by Equation \ref{eq:tacc}. Additionally, we also demonstrate that increasing the grain size leads to weaker acceleration due to the reduction in dust absorption cross-section. In contrast, within the highly optically thick regime ($\rm N_H \sim 10^{26} \, \text{cm}^{-2}$), denser gas columns experience a lower effective acceleration due to the absorption of UV flux by a thin inner shell, resulting in reduced momentum transfer from the outer shell. These findings align with the previous studies mentioned. To further support our observations, we calculate the $\rm t_{\rm diff}/t_{\rm acc}$ ratio in Equation \ref{eq:ratio}, which further validates the conclusions. However, it is important to acknowledge that our simulations may underestimate this effect since we did not account for photon downgrading, which has the potential to diminish the effectiveness of momentum transfer. Further, we find that the conditions in our simulations, which all result in successful outflows, also satisfy the outflow launching conditions outlined in the studies above. 

However, the conditions required for torus ejection may not apply to all AGNs, and our simulation represents only one particular scenario. Our results are specific to the assumptions of a massive black hole emitting at the Eddington limit, resulting in a high luminosity that ensures the ejection of dusty gas. This choice is made to ensure that the radiative acceleration is greater than the opposing gravity force and thus would result in the ejection of the dusty gas. As this condition would be maintained at larger distances from the AGN, if the AGN torus is successfully ejected, we expect it to escape the gravitational pull of the BH. Therefore, it is plausible that the outflow from the torus is part of an evolutionary sequence as suggested by observations \citep{banerji2012heavily, glikman2012first}. However, our current simulations only focus on a small region, therefore, we cannot provide a comprehensive analysis on this topic at this stage.

\begin{figure*}
    \centering
    \includegraphics[width=0.8\textwidth]{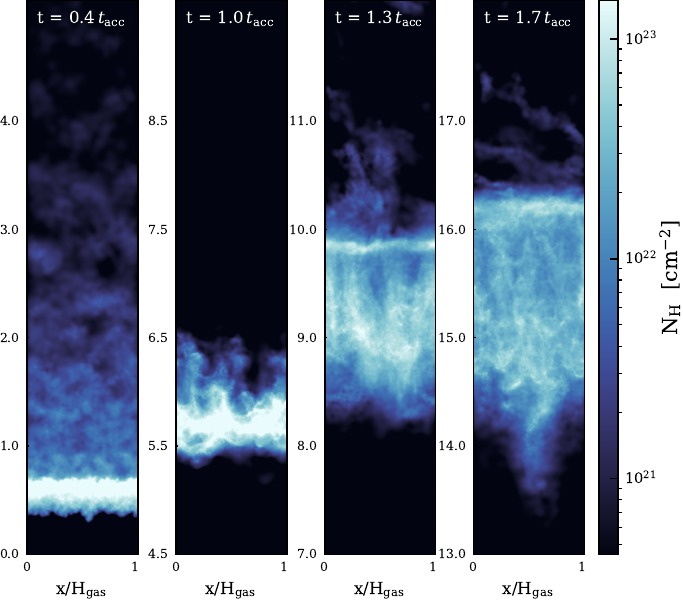}
    \caption[width=0.45\textwidth]{The evolution of the gas column density for an RDMHD simulation box with $\tilde c \sim  0.1 \, c$, $\rm N_H \sim 10^{24}cm^{-2}$ and $\grainsizemax \sim 0.01 \micron$ in the $xz$ plane at $t\sim (0.4, 1.0, 1.3, 1.7) \, t_{\rm acc}$, where for this case $t_{\rm acc} = \sqrt{20 \Lscale/ | a_{\rm eff}|}$. For this simulation, we initialize the box such that in the perfect dust-to-gas coupling limit, the net force from gravity is stronger than the radiation pressure force by a factor of $\sim 3$. The plot shows that despite the strength of gravity being stronger than the radiation-driven acceleration, a non-negligible component of the dust and gas is still ejected, however, the resulting ejecta is more compressed relative to our default setup, and a somewhat larger fraction ``falls back''. }
    \label{fig:failed}
\end{figure*}

We also study the behaviour of the wind in an environment where gravity dominates over the radiation-driven acceleration, i.e. where $\rm \tau_{\rm IR} \rm L/c \lesssim g M_{gas}$, or in our dimensionless units $\tilde \alpha/\tilde g \lesssim 1$ (though we note we are only modestly in this regime here, with gravity a factor of $\sim 3$ stronger than radiation). We show the projected morphology of the gas column under these conditions evolved to $t\sim (0.4, 1.0, 1.3, 1.7) \, t_{\rm acc}$ in Figure \ref{fig:failed}. As the net vertical acceleration is in the negative $\hat{z}$ direction, we define the acceleration time as $t_{\rm acc} = \sqrt{20 \Lscale/ | a_{\rm eff}|}$ for this simulation. The simulation is run with full RDMHD with the following parameters: $\tilde c \sim  0.1 \, c$, $\rm N_H \sim 10^{24}cm^{-2}$ and $\grainsizemax \sim 0.01 \micron$. Naively, we would expect a failed wind to result from these conditions, however, as shown in the plots, much of the gas (and dust as they are tightly coupled in this simulation), is successfully ejected. The increased strength of gravity does not cause the wind to halt, but rather compresses the gas and dust to a more compact ``shell''. After the ejecta is compressed into a thin slab, the radiation continues to accelerate the material, resulting in a thicker slab with prominent substructure at later times. Some gas indeed ``falls back'' - more than in our fiducial simulations with $\rm \tilde{g} < \tilde{\alpha}$; but the same in-homogeneity that allows tens of per cent of gas to ``fall down'' in those simulations leads to tens of per cent gas ejected here. 

When comparing the wind energetics from our simulations with the observations of AGN galactic outflows, such as those reported in \citet{fluetsch}, we find relatively consistent values of $\sim$50\% momentum loading within the wind relative to $\tau_{\rm IR} L/c$. For our fiducial AGN luminosity of $10^{46}$ erg/s, this translates to momentum rates in the range of $10^{35}-10^{37}$ g m/s$^{2}$ and kinetic rates in the range of ~$10^{43}-10^{45}  \rm$ erg/s. However, we must emphasize that our simulations are highly idealized and are based on several assumptions about the setup and thermodynamics of the outflow. For instance, our simulations do not account for the multi-phase structure of the gas or the processes that may alter energy dissipation, such as heating and cooling due to photoelectric and radiative processes such as line emission.

Additionally, the existence of polar dusty outflows in AGN has been suggested by recent interferometric observations \citep{asmus2016subarcsecond, honig2017dusty, alonso2021galaxy}. However, it is important to note that our current study is limited to a localized region within the obscuring torus, and that our simulations are agnostic to the overall geometry of the system. We explore different lines-of-sight and angles relative to the torus by varying the column density in our simulations. Specifically, the densest column density ($\rm N_H \sim 10^{26}$ cm$^{-2}$) corresponds to roughly equatorial lines-of-sight, while a column density of $\rm N_H \sim 10^{22}$ cm$^{-2}$ represents weakly obscured or more polar sight-lines. In Figure \ref{fig:tau}, we demonstrate that at $\rm N_H \sim 10^{22}$ cm$^{-2}$, our simulations still exhibit outflows. However, we would like to emphasize that this is expected because the simulations are set up such that radiation pressure on dust is stronger than the gravitational pull of the central source. It is important to acknowledge that our simulations treat all the physics consistently and assume the same dust composition throughout, without explicitly considering the properties of polar dust which could vary in composition and grain size  \cite[see][]{honig2017dusty,garcia2022torus, isbell2022dusty}. Regrettably, these factors are beyond the scope of our current study. However, we recognize the significance of investigating these additional factors, and in future work, we intend to conduct more comprehensive simulations that encompass the entire region surrounding AGN and account for the different dust properties.

\subsection{Gas and Dust Clumping and Coupling in AGN Winds
\label{sec:clumping}}
\begin{center}
\begin{figure*}
    \centering
    \includegraphics[width=\textwidth]{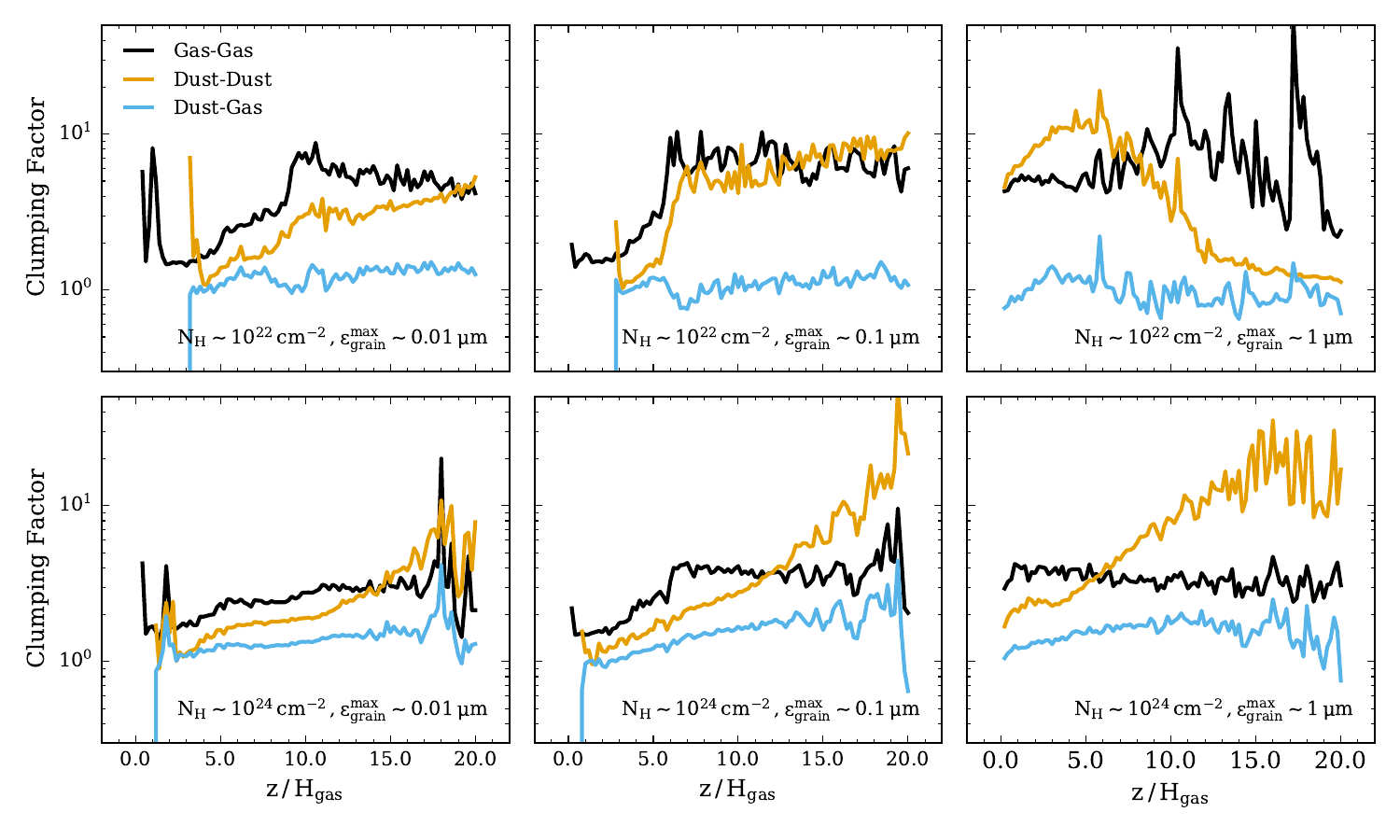}
    \vspace{-7mm}
    \caption{Clumping factors for gas-gas ($\langle \gasden^{2} \rangle / \langle \gasden \rangle^{2}$), dust-dust ($\langle \dustden^{2} \rangle/\langle \dustden \rangle^{2}$), and gas-dust ($\langle \gasden\,\dustden \rangle / \langle \gasden \rangle\,\langle \dustden \rangle$) at $\rm t\sim \text{min}(t_{\rm acc}, t_{\text{esc}})$. From left to right, the maximum grain size $\grainsizemax$ corresponds to $0.01 \micron$, $0.1 \micron$, and $1 \micron$ respectively, for an average column density of $10^{22}\text{cm}^{-2}$ (top) and $10^{24}\text{cm}^{-2}$ (bottom) within the simulation box. Gas-gas, dust-dust and gas-dust clumping is significant near the centre of the wind where most of the mass resides. Further, dust-dust and gas-dust clumping is stronger for larger grains. For an extended discussion, refer to \S \ref{sec:clumping}.}
    \label{fig:clumping}
\end{figure*}
\end{center}

As discussed above, we find that the dust and gas within the fluid are not always perfectly coupled. In Figure \ref{fig:clumping}, we quantify this by computing the gas-gas, dust-dust, and dust-gas clumping factors defined in Equation \ref{eq:clustering}, as a function of height within the simulation box at $\rm t\sim \text{min}(t_{\rm acc}, t_{\text{esc}})$, where $t_\text{esc}$ is the time at which 10\% of the dust/gas has escaped the top of the box.

\begin{align}
\nonumber C_{nm} &\equiv \frac{\langle \rho_{n}\,\rho_{m} \rangle_{V}}{\langle \rho_{n} \rangle_{V}\,\langle \rho_{m}\rangle_{V}} \\
&= \frac{V\,\int_{V} \rho_{n}({\bf x})\,\rho_{m}({\bf x})\,d^{3}{\bf x}}{\left[ \int_{V} \rho_{n}({\bf x})\, d^{3}{\bf x} \right] \, \left[ \int_{V} \rho_{m}({\bf x})\, d^{3}{\bf x} \right]} = 
\frac{\langle \rho_{n} \rangle_{M_{m}}}{\langle \rho_{n} \rangle_{V}}
\label{eq:clustering}
\end{align}

As shown in the equation, the clumping factor is analogous to the auto-correlation (for like species) and the cross-correlation (for different species) function of the local density field, where factors less than 1 imply an anti-correlation. We report clumping factors $\sim 1-10$ for the gas-gas and dust-dust clumping factors, and $\sim 1$ for dust-gas clumping. The gas-gas clumping factors, $\rm C_{gg}$, are lower at the base of the wind and increase up to a roughly constant value within the accelerated wind. As the gas is collisional and pressurized, its clumping is limited by pressure forces, especially on small spatial scales inside the wind. We note that for the run with $\rm N_{H} \sim 10^{22} \, cm^{-2}, \grainsizemax \sim 1 \micron$, the gas has high clumping factors at $\rm z\sim 10 \to 20 \: H_{gas}$. This occurs for this parameter space, due to the low gas column density and high acceleration forces, which make the gas effectively more compressible. Within this environment, the gas is subjected to intense radiation, resulting in strong acceleration forces acting upon it. Low density gas, characterized by higher compressibility, would experience larger relative fluctuations in density. These fluctuations give rise to localized density variations that exhibit strong correlations on small scales. As a consequence, the spatial density auto-correlation function reflects stronger correlations and higher clustering factors.

The dust-dust clumping factors, $\rm C_{dd}$, show a constant rise as a function of height to reach maximal values at the top of the box, and the slope of the profile weakly increases with grain size and weakly decreases with density. However, the run with $\rm N_{H} \sim 10^{22} \, cm^{-2}, \grainsizemax \sim 1 \micron$ shows a seemingly different behaviour as it corresponds to $t \sim t_{\rm esc}$. In this case, $t_{\rm esc}$ is smaller than $t_{\rm acc}$ due to poor fluid coupling under the specific conditions of the simulation. As a result, the dust distribution in the simulation shows more mass towards the bottom of the box, with a smaller amount of dust present at the top. The reason for this discrepancy is that the dust at the top has mostly escaped, while the majority of the dust remains concentrated at lower positions due to insufficient time to accelerate to higher positions. As a consequence, in this simulation, the clumping factors show an upward trend towards regions with higher dust density and decrease with height where there is less dust present.

In the general case, if we assume that $\rm C_{dd}$ is purely driven by the saturation of the RDIs, we expect clumping at some height $z$ to be stronger where the RDI growth time at a given wavelength is shorter. Plugging in equilibrium values of $\driftvel$ and $t_s$ in the super-sonic limit into Equation \ref{growthrate}, we obtain

\begin{align}
    \nonumber \rm t_{grow}(\lambda, z) &\sim \Bigg ( \frac{\lambda^4 \rhogas e^{-z/\Lscale} c_s^3}{a_{\rm eff} \internaldensity \grainsize (\dustgas)^5} \Bigg)^{1/6} \\
    &\propto \rhogas e^{-z/6\Lscale}. 
    \label{eq:tgrow_exp}
\end{align}
As all the parameters in the expression above except for the stratified density term are roughly independent of height, we expect the RDI growth timescale to get shorter as a function of height. In turn, the degree of dust clumping would increase as a function of height (clumping is $\sim 5$ times stronger for a factor $\sim 10$ increase in height) as shown in our plots. We note that this effect is suppressed for some of our simulations which could arise due to the non-linear evolution of the RDI's and/or competing processes such as turbulence. 

\begin{center}
\begin{figure*}
    \centering
    \includegraphics[width=\textwidth]{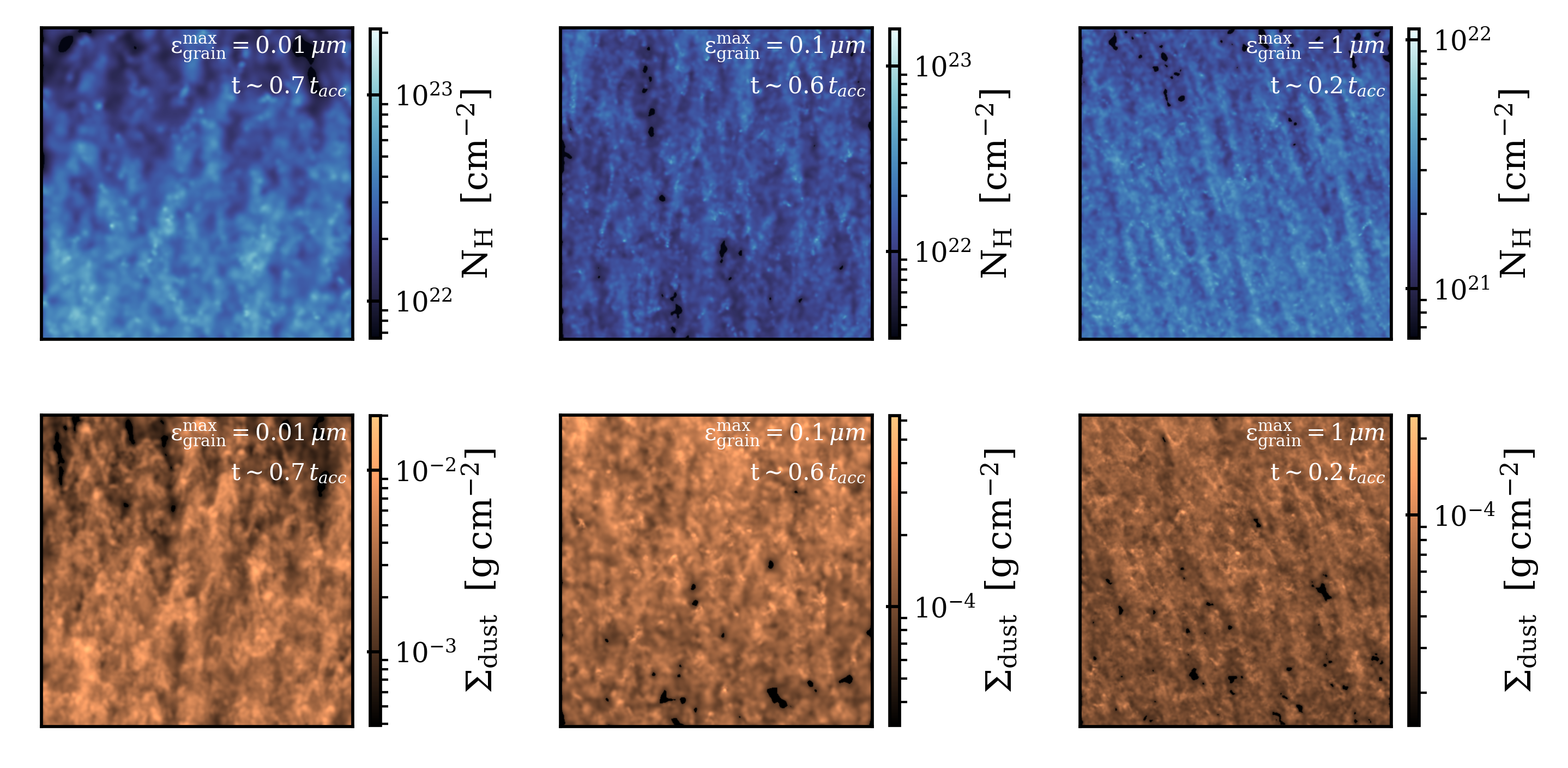}\hfil
    \caption[width=0.45\textwidth]{Gas column density (top) and dust surface density (bottom) within narrow bins in a zoomed-in region of high density within the AGN wind projected along the $xz$ plane at $\rm t\sim \text{min}(t_{\rm acc}, t_{\text{esc}})$ (where $\rm t_\text{esc}$ is the time at which 10\% of the dust/gas has escaped the top of the box). From left to right, the maximum grain size $\grainsizemax$ in the simulation box corresponds to $0.01 \micron$, $0.1 \micron$, and $1 \micron$ respectively, for an average column density of $10^{22}\text{cm}^{-2}$ at times 0.7, 0.6 and 0.2 $\rm t_{acc}$. Note that as the absorption efficiency is grain size dependent, the dust surface density is proportional to the extinction with $\rm A_\lambda \sim 0.1 (\Sigma/10^{-4} g \, cm^{-2}) (\micron / \grainsizemax)$. Larger grains show stronger clumping and thus more defined filaments.}
    \label{fig:zoombox_gas_g}
\end{figure*}
\end{center}

In Figure \ref{fig:zoombox_gas_g} we plot the zoomed-in column density profiles of the gas (top) and dust (bottom) in several RDMHD simulations. From left to right, the maximum grain size $\grainsizemax$ corresponds to $0.01 \micron$, $0.1 \micron$, and $1 \micron$ respectively, for an average column density of $10^{22}\text{cm}^{-2}$ within the simulation box. The structures formed appear more diffuse for smaller grain sizes. Usually, we see sharper structures for lower $\tau_{\rm geo}$, which could be shown by considering how $\rm t_{grow}$ depends on $\rm \tau_{geo}$. In Equation \ref{eq:time}, we showed that $\rm t_{acc}/t_{grow} \propto \tau_{geo}^{-1/6}$, therefore environments with lower $\rm \tau_{geo}$ would result in sharper structure. 

As the micro-scale structure of the dust within the torus is not spatially resolved observationally, we cannot directly compare the structures formed within our simulations to observations. Nonetheless, the physical variation in column densities could be inferred from the time variability for AGN sources. We discuss this in further detail in Subsection \ref{subsec:var}.

\subsection{Evolution of Velocity Fluctuations}

\begin{figure}
    \centering
    \includegraphics[width=0.45\textwidth]{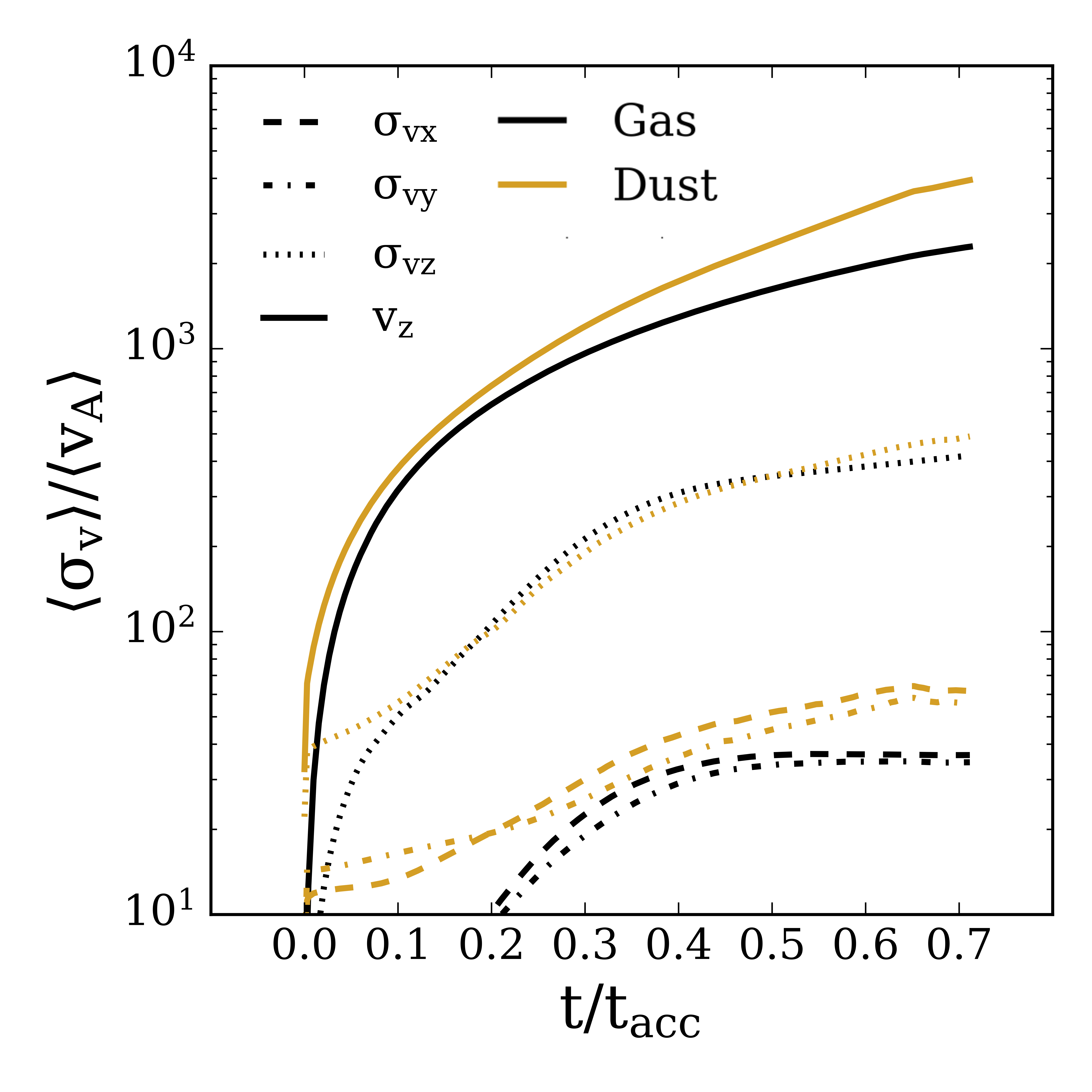}
    \caption{The temporal profile of the gas (black) and dust (yellow) velocity dispersion components ($\rm \sigma_{vx} \, ,\sigma_{vy} \, ,\sigma_{vz}$) and outflow velocity $\rm v_z$ relative to the box averaged \Alf\ speed ($\rm v_A$), for a simulation box with $\rm N_H \sim 10^{24} cm^{-2}$, $\grainsizemax \sim 1 \micron$. The RMS random velocity dispersion quickly saturates in all directions for both the gas and the dust. The RMS dispersion is dominated by the $\hat{z}$-component ($\sim 10\%$ variation), i.e. the direction of the outflow, due to slightly different drift speeds for the gas, different dust sizes and different sub-structures. The $\hat{x}$ and $\hat{y}$ components are $\sim 1$ order of magnitude weaker.}
    \label{fig:sat_time}
\end{figure}

To further analyze the evolution of the resultant non-uniform internal structure of the outflows within our simulations, we explore velocity fluctuations in dust and gas here. It is important to note that there are multiple RDI modes present simultaneously within the simulation box, and while the short wavelength modes will have the shortest growth times \citep{hopkins:2017.acoustic.RDI}, the dynamics will be dominated by the large-scale modes, as well as non-linear effects and in-homogeneity in the wind (eg. different clumps/ filaments moving differently).

\begin{figure}
    \centering
    \vspace{0.1in}
    \includegraphics[width=0.5\textwidth]{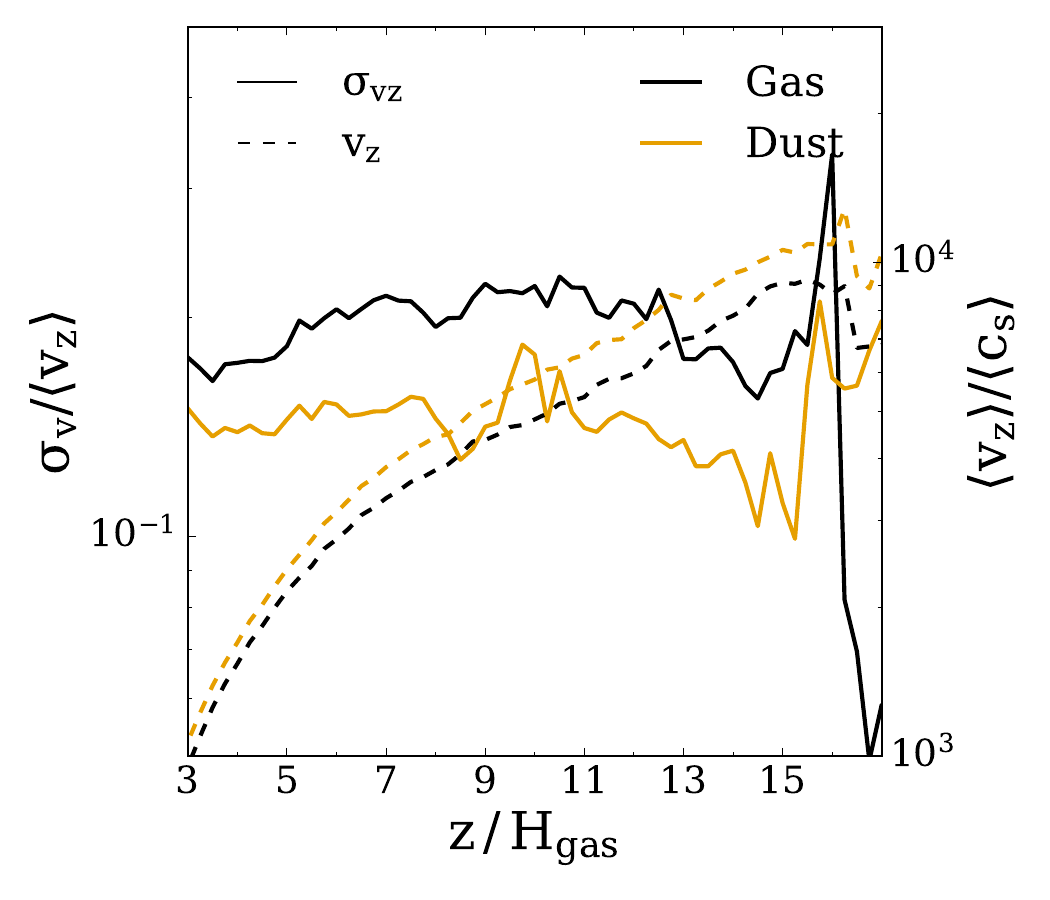}
    \vspace{-0.1in}
    \caption{The spatial profile of the gas and dust random velocity dispersion (RMS) in the $\hat{z}$ direction normalized to the average outflow velocity, $\langle \rm v_z \rangle$ at a given height $z$, for a simulation box with $\rm N_H \sim 10^{24} \, cm^{-2}$, $\grainsizemax \sim 1 \micron$ at $\rm t \sim 0.7 \, t_{acc}$.We also present the average flow velocity, $\rm \langle v_z \rangle$, normalized to the average sound speed within the box, $\langle \cs \rangle$. The plot illustrates that the fluid (gas + dust) reaches highly supersonic velocities (approximately $4 \times 10^4 , \rm km , s^{-1}$) and that the ratio of $\rm \sigma_{v_z}/v_z$ remains relatively constant as a function of height within the box. We only show the $\hat{z}$-component of the dispersion in this plot as the $\hat{x}$ and $\hat{y}$ components show a similar behaviour but a magnitude weaker in the ratio of their respective velocity dispersion to the outflow velocity.}
    \label{fig:sat_space}
\end{figure}

Figures \ref{fig:sat_time} and \ref{fig:sat_space} show the evolution of gas and dust turbulent velocity components. Figure \ref{fig:sat_time} displays the normalized root mean squared (RMS) random velocity dispersion for $\hat{x}$, $\hat{y}$, $\hat{z}$, and 3D components over time. Figure \ref{fig:sat_space} illustrates the normalized RMS velocity dispersion in the $\hat{z}$ direction, $\rm \sigma_{vz}$, and mean outflow velocity, $\rm v_{z}$, as a function of height. The plot shows the behaviour for our $\rm N_H\sim 10^{24}cm^{-2}, \grainsizemax \sim 1 \micron$ run, however, we note that we observe the same behaviour throughout our parameter space. The dispersions grow exponentially fast (as expected if they are RDI-driven) at early stages and quickly saturate (within 0.1-0.2 $\rm t_{acc}$) for all runs within our parameter set. This suggests that in an AGN tori, such instabilities have already saturated within the time taken to launch a wind, and later structure formation is mostly driven by radiation-pressure accelerating the medium in addition to the turbulence within the flow.  

Further, at the non-linear stage of their evolution, the gas and dust both reach similar super-\Alf{ic} random velocities with the RMS dispersion dominated by the $\hat{z}$-component. The $\hat{x}$ and $\hat{y}$ components are $\sim 1$ order of magnitude weaker due to the inherent geometry of the problem and the relatively weak Lorentz forces (i.e. $\rm v \sim v_z \gg v_A$). As the turbulence is super-\Alf{ic}, the magnetic field has a weak influence on the flow dynamics, resulting in isotropic turbulence in the $\hat{x}$ and $\hat{y}$ directions as the magnetic field does not introduce significant anisotropy.

Analysing the spatial profile, we note $\sim 20\%$ and $\sim 2\%$ dispersion in the $\hat{z}$ and $\hat{x}-\hat{y}$ directions respectively relative to the outflow velocity. Towards the base ($z \sim 0-3 \, \Lscale$), and top ($z \sim 17-20\, \Lscale$) of the wind, the profile shows anomalous behaviour due to the presence of a relatively low number of dust particles/gas cells and boundary effects. Away from the boundaries, the dispersion shows no spatial dependence. In addition, we present the spatial profile of the outflow velocity, $\rm v_z$, shown as a dashed line. We observe a consistent trend of increasing outflow velocity with height, as particles with higher velocities can travel further in a given time interval. Further, we note that the outflow attains highly super-sonic velocities. By assuming an isothermal sound speed and a range of molecular gas temperatures between $\rm T \sim 10^3-10^4 \, K$, we estimate that corresponds to a maximum outflow velocity range of approximately $2-6 \times 10^4 \rm km s^{-1}$. We compared our estimates with the observed velocities reported in \citet{fiore2017agn} for AGNs with similar luminosities and find them to be consistent with X-ray winds with ultra-fast outflows. Therefore, while the comparison provides some insights, the velocities we observe in our simulations may not be directly comparable as they likely originate from different physical mechanisms and/or locations. However, as AGN outflows can arise from various physical processes, there are likely multiple mechanisms driving the observed outflows. Therefore, we caution against drawing definitive conclusions based solely on this comparison.

\section{Predicted AGN Variability}
\subsection{Temporal and Spatial Variability in Column Densities along Observed sight-lines}
\label{subsec:var}

\begin{center}
\begin{figure*}
    \centering
    \includegraphics[width=\textwidth]{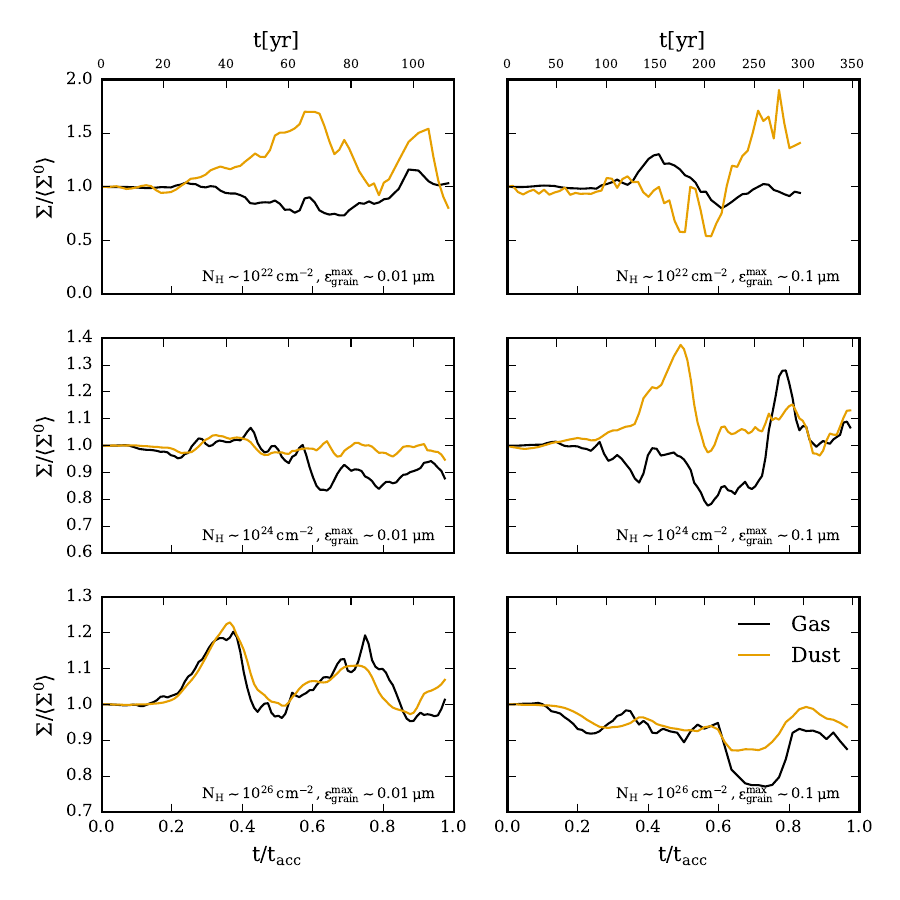}
    \vspace{-5mm}
    \caption{The sight-line integrated surface density $\Sigma$ along a random line-of-sight towards the AGN accretion disk. We normalize $\Sigma$ to $\Sigma^{0}$, the initial mean surface density in the simulation box for convenience. We compare both gas and dust columns, from top to bottom, the total gas column density $\rm N_{H}$ in the simulation box corresponds to $10^{22}\text{cm}^{-2}$, $10^{24}\text{cm}^{-2}$, and $10^{26}\text{cm}^{-2}$ respectively, and maximum grain size is $0.01 \micron$ (left), $0.1 \micron$ (right). Overall, the dust and gas show fluctuations of similar amplitude, and there is an order of a few \% variability on short timescales (a few years), with higher variation ($\sim 10-40\%$) on long timescales relative to the acceleration time of the wind. However, at a given time the gas and dust $\Sigma$ fluctuations do not exactly match. }
    \label{fig:nh}
\end{figure*}
\end{center}

\begin{center}
\begin{figure*}
    \centering
    \includegraphics[width=\textwidth]{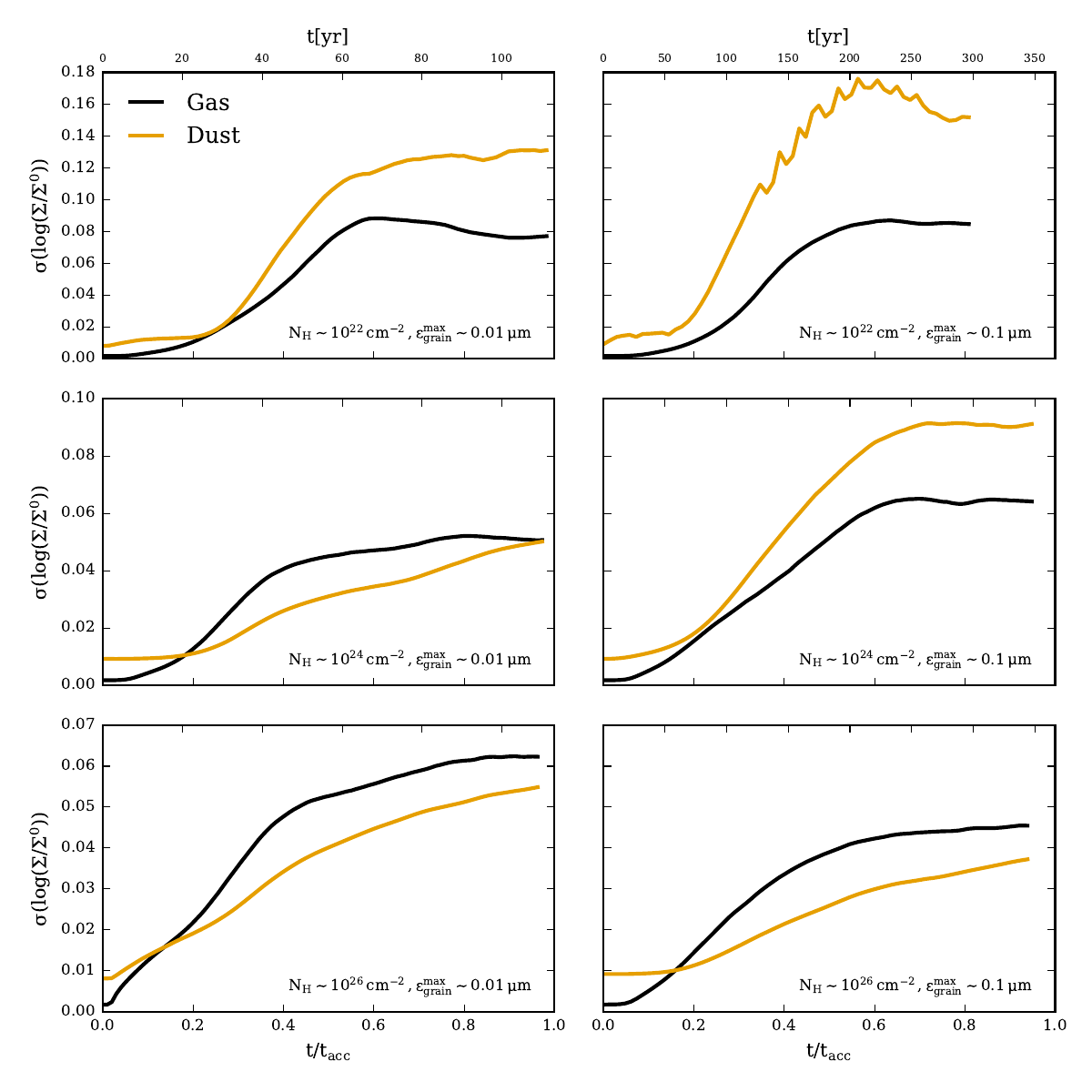} 
    \vspace{-5mm}
    \caption{The sight-line-to-sight-line spatial variability of the gas and dust integrated surface densities across different sight-lines within the box as a function of time. We specifically plot $1 \sigma$ dispersion in the log of the surface density compared across 100 random sight-lines to the AGN accretion disk, through the wind, at each time t. From top to bottom, the total gas column density $\rm N_H$ in the simulation box corresponds to $10^{22}\text{cm}^{-2}$, $10^{24}\text{cm}^{-2}$, and $10^{26}\text{cm}^{-2}$ respectively, and maximum grain size is $0.01 \micron$ (left), and $0.1 \micron$ (right). Both the gas and the dust show similar degrees of variability, with the dust variability increasing at a higher rate at later times. We note that below $\rm N_H \sim 10^{26} cm^{-2}$, larger grains result in a larger variation due to more prominent vertical filaments across the simulation, however, the grain size has a minimal effect on the spread of the distribution for higher column densities. }
    \label{fig:nh_vartime}
\end{figure*}
\end{center}

\begin{center}
\begin{figure*}
    \centering
    \includegraphics[width=\textwidth]{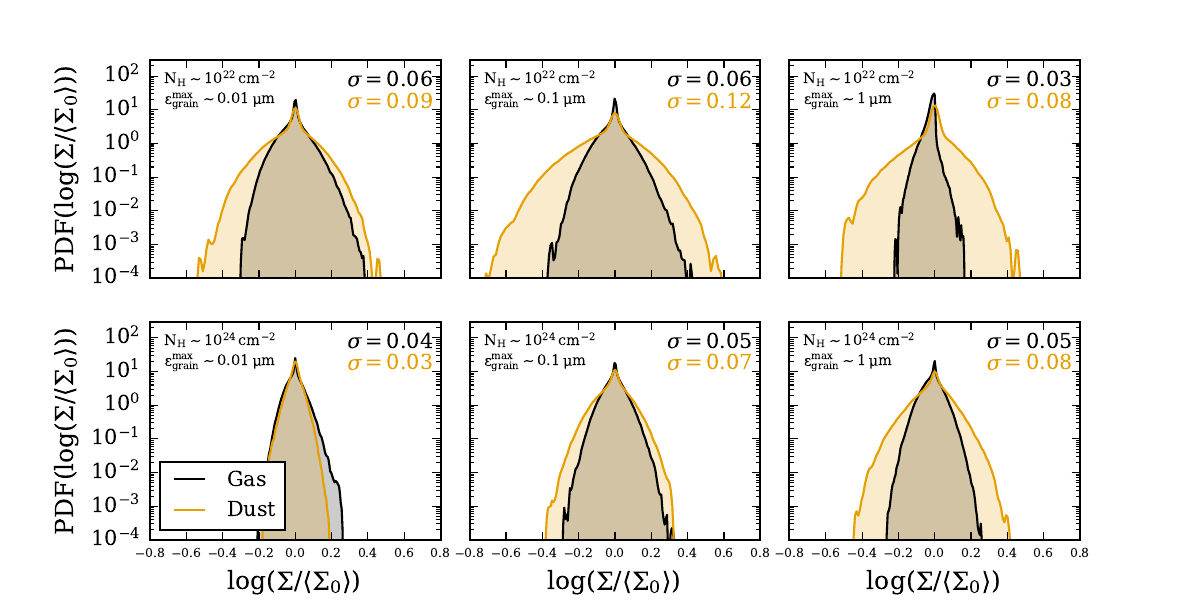}
    \caption{The normalized probability density function (PDF) of the surface density for the dust and gas components across 100 random sight-lines at each time, combining all times after the wind begins to launch ($\rm t > 0.1 t_{acc}$). From top to bottom, the total gas column density $\rm N_\text{H}$ in the simulation box corresponds to $10^{22}\text{cm}^{-2}$ and $10^{24}\text{cm}^{-2}$, and maximum grain size is $0.01 \micron$ (left), $0.1 \micron$ (middle), $1 \micron$ (right). The PDFs show distributions that are highly non-gaussian with a narrow peaked core component and wings with steep drop-offs as a result of enhanced fluctuations. On average, the dust shows a higher spread in the distribution than the gas as expected given its collisionless nature. This difference in spread decreases as the fluid approaches the limit of being perfectly coupled, i.e. smaller grain sizes, and higher column densities, again as expected for the RDI's.}
    \label{fig:nh_varpdf}
\end{figure*}
\end{center}

While it is difficult to resolve the underlying structure of the dust within AGN tori, AGN spectra and SEDs with high temporal resolution can be obtained which could probe these small scale fluctuations. The methodology employed here closely follows that presented in \citet{steinwandel:2021.dust.rdi.variable.stars} to which we refer for details. In Figure \ref{fig:nh}, we compute the time variability in the sight-line integrated surface density ($\Sigma$) of the dust and gas integrated for an infinitesimally narrow line-of-sight down the $\hat{z}$ direction i.e., towards the accretion disk which should have an angular size that is very small compared to our resolution (hence an effectively infinitesimally narrow sight-line), and show the variance of the distribution in Figure \ref{fig:nh_vartime}. From top to bottom, the total gas column density $\rm N_{H}$ in the simulation box corresponds to $10^{22}\text{cm}^{-2}$, $10^{24}\text{cm}^{-2}$, and $10^{26}\text{cm}^{-2}$ respectively, and maximum grain size is $0.01 \micron$ (left), and $0.1 \micron$ (right). The plots show variability of order a few \% for both the dust and gas over relatively short timescales (a few years in physical units), and up to $\sim 20-60 \%$ variation on long timescales (decades). The amplitude of the short-timescale variations is roughly independent of maximum grain size and decreases for denser columns.

This result is consistent with our findings for the underlying morphological structure of the wind, where for low column density boxes with large grains, we find that the RDIs drive the formation of defined dense vertical filaments which would cause significant variability as they cross the line-of-sight. Further, the variability extends beyond the time taken for the instabilities to grow and is likely driven by the large velocity dispersion of the dust and gas. However, while the magnitude of the velocity dispersion is similar across all our runs, denser columns form more randomized clumps which are likely to be averaged over when integrating down the $\hat{z}$ direction and thus result in weaker variation in the sight-line quantities compared to the 3D quantities \citep[see e.g. ][]{hopkins:2021.dusty.winds.gmcs.rdis}.

In principle, fluctuations in the integrated surface density could also exhibit corresponding fluctuations in the line-of-sight grain size distribution, as shown in the case of AGB-star outflows studied in \citet{steinwandel:2021.dust.rdi.variable.stars}. Therefore, we analyze the spatial fluctuations in the grain size distribution in the same manner as $\Sigma$. However, we find that the fluctuations in the grain size distribution are significantly weaker than the environments studied in \citet{steinwandel:2021.dust.rdi.variable.stars} (perhaps consistent with our $\Sigma$ fluctuations themselves being much weaker), and largely fall within the range we might expect from shot noise given our limited resolution (the shot noise being large for grain size fluctuations since we must consider only a narrow range of grain sizes, hence a more limited number of dust particles). Therefore we cannot conclusively say whether or not there is a potentially measurable correlation between the fluctuations in $\Sigma$ and the grain size distribution.

In Figure \ref{fig:nh_vartime}, we show the spatial variability of the logarithmic gas and dust integrated surface densities computed over all possible sight-lines as a function of time. As one would expect, the variability in surface density increases as the RDIs develop. For lower column densities, the dust surface density shows higher variability for larger grain sizes. However, for simulations with column densities $\rm N_H \gtrsim 10^{22} cm^{-2}$, we observe a weak dependence of the surface density variation on the grain size, where the dust and gas exhibit similar levels of variation across different maximum grain sizes. Further, we note a trend of decreasing variation for increasing column densities. We are currently unaware of any significant observational constraints related to this particular trend. The lack of constraints can be attributed to the high column densities (Compton thick) found in these environments and the predicted long timescales on which variability occurs, spanning from decades to hundreds of years. Consequently, studying Compton thick sources presents significant challenges. The anticipated variability in these sources is unlikely to be detected within the X-ray band, but it may manifest as a modulation of UV/IR radiation due to dust. Although this effect has not been dis-proven by observations, it is crucial to consider other factors, such as detailed cooling and heating physics, that could drive further variability within this regime. This underscores the need for further research to determine the primary sources of variability in Compton thick environments. Therefore, for column densities of this magnitude, it is plausible that RDIs may not be the primary driver of variability.

To interpret the trend in the variability, we follow the analysis presented in \citet{moseley:2018.acoustic.rdi.sims}. Assuming pure isothermal MHD turbulence, the variance of the gas density field will roughly follow a log-normal distribution of the form, 
\begin{align}
    \rm \sigma^2(ln(\rhogas)) = ln(1+(b |\sigma_v/c_s|)^2),
\end{align}
where b corresponds to the ''compressibility'' of the fluid with $b \sim 0.2-1$. We expect the saturation amplitude of the turbulence within the box to occur when the eddy turnover timescale is of order the growth timescale of the instability mode, this results in the following scaling for the long-wavelength regime,

\begin{align}
    \sigma_v \sim (\dustgas)^{1/3}(k \langle c_s\rangle \langle t_s \rangle)^{2/3} (\langle \driftvel \rangle/c_s)^{2/3}.
\end{align}
Therefore, by combining both relations, we get, 
\begin{align}
\rm
\sigma(log(\rhogas)) \sim \ln(1+ \rhogas^{-4/3} \grainsize^{1/3}).
\end{align}

So in the case where $\rhogas^{-4/3} \grainsize^{1/3} \gg 1$, the variability will be higher for columns with lower density and larger grain sizes with a strong dependence on the density and a weak dependence on the grain size. However, when $\rhogas^{-4/3} \grainsize^{1/3} \ll 1$, the variability will be roughly similar at all densities and grain sizes. 

In Figure \ref{fig:nh_varpdf}, we show the normalized PDF of the logarithmic surface density field for all times after the saturation of the RDIs and all sight-lines. From top to bottom, the total gas column density $N_\text{gas}$ in the simulation box corresponds to $10^{22}\text{cm}^{-2}$ and $10^{24}\text{cm}^{-2}$ respectively, and maximum grain size is $0.01 \micron$ (left), $0.1 \micron$ (middle), $1 \micron$ (right). We omit the plots for larger column densities but report that they are similar to the bottom left plot. As shown in the plots, the profile of the PDFs is highly non-gaussian with a narrow peaked core component and wings that sharply drop off, indicative of strongly enhanced fluctuations. The dust PDF is broader than that of the gas at lower column densities and higher grain sizes, i.e. when the dust is not well-coupled with the gas. This difference is negligible for more obscured lines-of-sight ($\rm N_{H} \gtrsim 10^{24} \, cm^{-2}$), as the fluid is strongly coupled across the range of grain sizes we consider.

\subsubsection{Power Spectral Analysis}
\label{subsec:power}
\begin{center}
\begin{figure}
    \centering
    \includegraphics[width=0.45\textwidth]{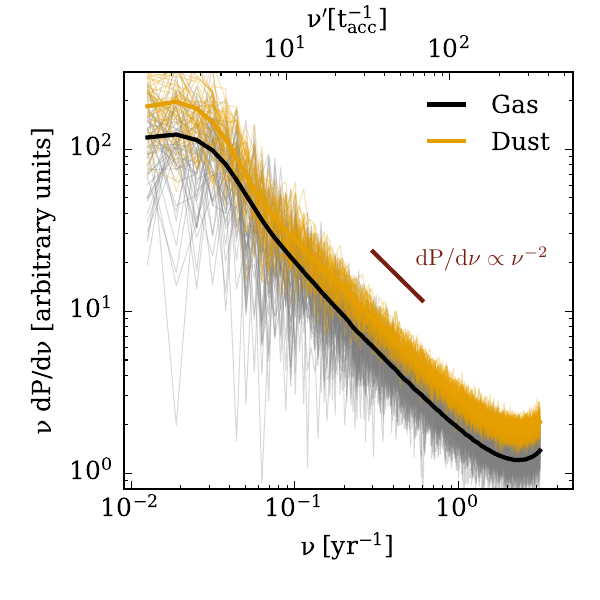}
    \caption{Temporal power spectrum of the gas and dust sight-line integrated surface densities along individual sight-lines as \fref{fig:nh}. Thick lines show the average over all sight-lines. This is for one RDMHD simulation with initial $\rm N_H\sim 10^{24} \, cm^{-2}$ and $\grainsizemax \sim 0.01 \micron$, but others are qualitatively similar. Both the dust and gas show similar profiles, with the dust carrying roughly twice the amount of power as the gas. The spectra show power with an approximate red-noise spectrum, $\rm dP/d \nu \propto \nu^{-2}$ over most of the resolvable time range. We expect that the power on long timescales is mostly driven by global processes such as the vertical acceleration of the fluid and that the power on shorter timescales is driven by the density fluctuations in the wind. }
    \label{fig:power_time}
\end{figure}
\end{center}

In Figure \ref{fig:power_time}, we present the temporal power spectrum, for individual lines-of-sight (as \fref{fig:fullbox_gas_t}) and averaged over all lines-of-sight, of the integrated gas and dust surface density in black and yellow thick lines respectively. We show this for a simulation with $\rm N_H\sim 10^{24} \, cm^{-2}$ and $\grainsizemax \sim 0.01 \micron$. We omit the spectra for the remainder of our simulation set as they show a similar profile. The spectra for dust and gas show similar profiles, with twice the amount of power present in the dust spectrum relative to the gas (consistent with our previous analysis). The plot indicates that most of the power is on long timescales, with a spectral index $\alpha_\nu \sim -2$, defined as $\rm dP/d \nu \propto \nu^{\alpha_\nu}$. This spectral index is very close to canonical red noise which is consistent with AGN observations probing comparable timescales\citep{caplar2017optical,macleod2012description}, and could arise from an array of physical processes. For instance, if we assume that on small scales, the density fluctuations take the form of a Gaussian random field, as the surface density is an integral over that field, it is natural that the resulting power spectrum would take this form. However, it is worth noting that observations of optical power spectral densities have indicated a range of slopes, with values often steeper than the canonical -2 value at high frequencies \citep{smith2018kepler, simm2016pan}.

Further, we note a break at the low-frequency end of the power spectrum, which corresponds to the acceleration timescale of the fluid within the simulation box. Similar breaks have been observed in AGN power spectra, which were found to be correlated with intrinsic properties of AGNs such as their mass \citep{burke2021characteristic}. However, these breaks were observed to occur on different timescales compared to the breaks in our simulations. The high-frequency plateau in our PSD, however, is likely an artefact  due to our limited time resolution and simulation duration. Overall, we acknowledge the complexity and variability of AGN power spectra and caution readers about the limitations of our simulations in capturing the full range of observed power spectrum behaviours.

\begin{center}
\begin{figure}
    \centering
    \includegraphics[width=0.45\textwidth]{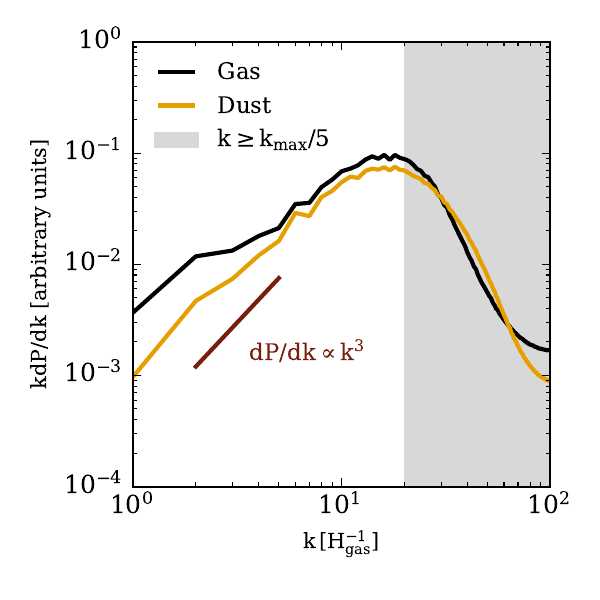}
    \caption{Spatial power spectrum of the three-dimensional dust and gas log density fields (log($\rm  \rho_{gas}/<\rho_{gas}>$)), (log($\rm \rho_{dust}/<\rho_{dust}>$)). We show this for one simulation with $\rm N_H\sim 10^{24} \, cm^{-2}$ and $\grainsizemax \sim 0.01 \micron$ at $\rm t \sim t_{acc}$, but others are qualitatively similar. The plot shows that similar power for the gas and dust that increases on small scales roughly according to $\rm dP/dk \propto k^{3}$ until the resolution limit is approached ($\rm k_{max}$ corresponds to the simulation resolution limit), after which due to numerical effects, power on smaller length scales decreases as modes are unresolved. }
    \label{fig:power_space}
\end{figure}
\end{center}

In Figure \ref{fig:power_space}, we show the spatial power spectrum of the logarithm of the 3-dimensional density field for a column with $\rm N_H\sim 10^{24} \, cm^{-2}$ and $\grainsizemax \sim 0.01 \micron$, and similar to above, note that it is roughly consistent with the spectra for our other simulations. The plot shows similar profiles for the dust and the gas, which is indicative that on the relatively large scales that we are probing, the dust and gas fluctuations are order-of-magnitude comparable. Further, the power increases exponentially with a spectral index $\alpha_k \sim 3$, defined as $\rm dP/dk \propto k^{\alpha_k}$ until a few factors of the resolution limit is hit, after which power on smaller length scales would not be resolvable. Thus the power decay on relatively short-length scales should be regarded as a numerical effect, as we expect it to continue to rise for smaller length scales.

%%%%%%%%%%%%%%%%%%%%%%%%%%%%%%%%%%%%%%%%%%%%%%%%%%

\subsubsection{Relation to AGN Observations}
The RDIs and other instabilities provide a natural explanation for the clumpy nature of the dusty torus, which together with the turbulent dynamics of the fluid, results in variability in the observed luminosity. While the variation we deduce is relatively small, it is non-negligible. We resolve $\sim 10 - 30 \%$ variation on scales of a few years which would be observable on human timescales. For Compton thick sources, such variability in the gas column would be detectable and significantly change the hardness of the observed X-rays and reduce luminosity by factors of $\sim 2$. However, the typical behaviour in our simulations would not give rise to variability similar to more extreme changing-look AGN which presumably is due to other physics (e.g. accretion disk state changes). 

We compare our results to optical variability studies by \citet{stone2022optical, macleod2010modeling, suberlak2021improving}. These studies report similar PSD slopes of -2, consistent absolute magnitude variability amplitudes, and characteristic break timescales on the order of years. While our simulations predict variability that extends to longer timescales and longer break timescales, determining such timescales would require longer observational baselines. Additionally, we take note of the X-ray variability observations by \citet{gonzalez2012x}, which also exhibit consistency with red noise characteristics. However, we acknowledge that X-ray variability is likely dominated by processes occurring in the accretion disk and operates on much shorter timescales. Therefore, while there is a similarity in the power spectrum slopes, it may not be the most suitable comparison for our simulations. While our model matches the reported PSDs in shape and magnitude, caution is advised in overgeneralizing the agreement. Red noise spectra can stem from widespread Gaussian processes, suggesting that other mechanisms are likely contributing to the observed variability.

 % that is consistent with a large array of AGN observations \citep{2002ApJ...571..234R,2005ApJ...623L..93R, markowitz2014first, laha2020variable, de2007broad

For our model, we expect the primary source of obscuration at optical/UV wavelengths to be the dust component, while at shorter wavelengths such as X-rays, we anticipate that gas will dominate the obscuration. A unique feature of our model is that it predicts a correlation between the variability at different wavelengths, with RDIs driving simultaneous variability at varying magnitudes depending on the observation wavelength. Therefore, the extinction at a given wavelength, $A_\lambda$, would be proportional to the dust surface density, $\delta \Sigma_{\text{dust}}$, and related by the extinction coefficient, $K_\lambda$, i.e., $\delta A_\lambda \sim K_\lambda \cdot \delta \Sigma_{\text{dust}}$. Based on previous estimates by \cite{draine:2003.dust.review}, we expect values of $K_\lambda$ to be around 5-10 in the optical band and 0.5-5 in the IR band. However, our simulations do not include the region interior to the sublimation radius often associated with AGN X-ray variability \citep{merloni2014incidence, middei2017long}. As this region is dust-free, any variability attributed to that region cannot be driven by the RDIs.

For our simulations, we predict several distinctive features that differentiate them from other models. One such feature is the relative variation between the dust and gas components. We observe fluctuations in the line-of-sight integrated dust-to-gas ratio, where the dust component varies independently of the gas component and sometimes in opposite directions. Observationally, this would manifest as instances where the UV spectrum becomes highly reddened due to increased dust obscuration while the X-ray spectrum remains relatively constant, or vice versa. Additionally, variations in the dust-to-gas ratio would introduce variability in the observed extinction curve.
Similar variability has been reported by \citet{10.1093/mnras/stad1774} which reports variability on decade timescales in the near-infrared (NIR) that does not correlate with the observed variability in X-ray gas reported by \citet{sanfrutos2016eclipsing} in regions corresponding to the dusty torus. Furthermore, there have been observations of sources where the X-ray flux varies by approximately 20\% to 80\% over a few years, with no apparent variation in the optical component \citep{2002ApJ...571..234R,2005ApJ...623L..93R, markowitz2014first, laha2020variable, de2007broad}.
Another feature predicted by our RDI simulations is the presence of high-velocity outflows that surpass the Keplerian velocity of the region. Observations by \citet{choi2022physical} in the AGN torus region have reported broad absorption lines corresponding to torus-like distances from the AGN source, indicating the presence of such high-velocity outflows that align with the predictions from our RDI model. 
However, if other mechanisms drive similar changes in the dust-to-gas ratio or high-velocity outflows, the observed variations may become degenerate, making it challenging to attribute the variability solely to the RDI mechanism.

In addition to that, we caution that our findings are sensitive to both the physical size of the line-of-sight/spatial resolution and the temporal resolution of our simulations. When considering observations, the thickness of the line-of-sight is limited by the size of the emitting region, i.e. the angular size of the AGN disk. Therefore to validate our choice of an infinitesimally narrow line-of-sight for our calculations, we consider the size of the AGN disk relative to the size of the torus. An AGN of luminosity $\rm 10^{46} erg/s$ with a disk emitting black-body radiation peaking in the near-UV regime with an effective temperature of $\rm \sim 10^4 K$ will have a radius, $\rm R_{d}$ of order $\rm R_d \sim \sqrt{L/4 \pi \sigma_{SB} T^4} \sim 3 \times 10^{-2} pc$, where $\rm \sigma_{SB}$ is the Stefan-Boltzmann constant. Therefore for torus of radius $\sim \rm 1.1 pc$, an infinitesimally narrow line-of-sight would be a reasonable approximation to an observationally limited line-of-sight. However, there have been cases where the continuum emission region has been resolved in the UV/IR waveband \citep{leighly2019z}. 

Regarding the timescales of the variability predicted by our analysis, we note that the shortest timescales we can resolve are limited by the frequency at which we output our snapshots ($\sim$ years), therefore we are not resolving variability on all human observable timescales and would expect that there would still be variability due to the RDIs on shorter timescales than those reported in this work. In addition, we expect that the variability that arises due to the RDIs would be much faster than that predicted by an occultation model.

\section{Conclusions}
\label{sec:conc}
In this work, we present simulations of radiation-dust-driven outflows explicitly accounting for dust dynamics and dust-gas radiation-magnetic field interactions, with initial conditions resembling AGN tori. We model the dust using a realistic grain size spectrum and grain charge under the influence of a radiation field and accounting for drag and Lorentz forces. The dust interacts with gas through collisional (drag) and electromagnetic (Lorentz, Coulomb) forces, which couple the two fluids and absorbs radiation which accelerates grains, determining whether they, in turn, can accelerate gas. While within this environment, the dust and gas are closely coupled in the sense that the ``free streaming length'' of dust grains is very small, explicit treatment of dust dynamics reveals that the fluid is unstable on all length scales to a broad spectrum of fast-growing instabilities. We summarize our key findings below. 

\begin{enumerate}[i]
    \item \textbf{RDIs:} The RDIs develop rapidly on scales up to the box size, forming vertical filamentary structures that reach saturation quickly relative to global timescales. We find that the behaviour of the RDIs is sensitive to the geometrical optical depth $\rm \tau_{geo}$ with environments with higher optical depths resulting in a more tightly coupled dust-gas fluid ($\rm \ell_{stream, dust}/\Lscale \propto \tau_{geo}^{-1}$ as shown in Equation \ref{eq:streaming}), and longer RDI growth times ($\rm t_{grow}/t_{acc} \propto \tau_{geo}^{1/6}$ as shown in Equation \ref{eq:time}). Other parameters such as AGN luminosity, gravity, grain charging mechanism, and the gaseous equation of state show weaker effects on the dynamics or morphology of the RDIs. 
    
    \item \textbf{Clustering:} The RDIs drive strong dust-dust and gas-gas clustering of similar magnitude (order of magnitude fluctuations) on small scales for all conditions explored within our parameter set. Thus, the RDIs provide yet another (of many) natural mechanism for explaining the clumpy nature of AGN tori. 

    \item \textbf{Outflows:} Our results show that both the dust and gas are accelerated to highly super-sonic velocities resulting in a wind which can successfully eject $70-90 \%$ of the gas present. In addition, the RDIs drive super-\Alf{ic} velocity dispersion of order $\sim 10\%$ of the outflow velocity. Further, while the morphological structure of the RDIs generates low opacity channels through which photons can in principle escape, we find that this ``leakage'' is modest, usually resulting in less than a factor of $\sim 3$ loss of photon momentum relative to the ideal case. In every case, the remaining momentum (for quasar-like conditions modelled here) is more than sufficient to drive a wind.
    
    \item \textbf{Integrated Surface Density Variation:} The resulting morphology and turbulence give rise to both short ($\leq$ years) and long timescale (10-100 years) variability in the column density of gas and surface density/opacity of dust integrated along mock observed lines-of-sight to the quasar accretion disk. These fluctuations have RMS amplitude along a given sight-line of order $\sim 10-20\%$ over year to decade timescales with a red noise power spectrum, consistent with a wide array of AGN observations. We note that both the dust and gas show variability on similar timescales that roughly follow similar trends statistically, but do not match 1-to-1 at any given time — they fluctuate relative to one another, providing a natural explanation for systems where dust extinction is observed to vary in the optical/NIR independent of the gas-dominated x-ray obscuration and vice versa \citep{2002ApJ...571..234R,2005ApJ...623L..93R, markowitz2014first, laha2020variable, de2007broad, smith2007x}. Our model suggests that the variability in the optical/NIR bands will be correlated in time and proportional to the variability in dust surface density. The X-ray variability, which is associated with the gas surface density variation caused by RDIs, is not expected to be strongly correlated with the optical/NIR variability.
\end{enumerate}
\acknowledgments{Support for for NS and PFH was provided by NSF Research Grants 1911233, 20009234, 2108318, NSF CAREER grant 1455342, NASA grants 80NSSC18K0562, HST-AR-15800. Numerical calculations were run on the Caltech compute cluster ``Wheeler,'' allocations AST21010 and AST20016 supported by the NSF and TACC, and NASA HEC SMD-16-7592.}

\datastatement{The data supporting this article are available on reasonable request to the corresponding author.}
\bibliography{example}{}
\bibliographystyle{mnras}

% Don't change these lines
% \bsp% typesetting comment
\label{lastpage}
\begin{appendix}
\section{List of Simulations}
\label{appendix:sims}
\begin{center}
\begin{table*}
\begin{tabular}{ l r r r r r r r r } 
 \hline
%  \hline
Name & $\rm N_{\rm H}$[$\rm cm^{-2}$]&$\epsilon_\text{grain}^{\text{max}} [$\micron$]$&$\sizeparammax$&$\chargeparammax $&$\gravparam$[$10^5$]&$\accparam$[$10^9$] & $\beta^0$ & Notes\\ 
\hline
n1e22\_eps1 &  1e+22& 1 & 6.5e-3 & 27.0 & 5.7 & 0.11 & 0.13& default\\
n1e22\_eps0.1 & & 0.1 & 6.5e-4 & 270.0 & 5.7 & 1.1 & 0.13& 10x smaller $\epsilon_\text{grain}^{\text{max}}$\\ 
n1e22\_eps0.01 &  & 0.01 & 6.5e-5 & 2700.0 & 5.7 & 11.0 & 0.13 & 100x smaller $\epsilon_\text{grain}^{\text{max}}$\\ 
\hdashline
n1e22\_eps1\_rhd & & 1 & 1e-2 & 10 & 1 & 0.05 & 1& default RDMHD\\
n1e22\_eps1\_rhd\_c & & 1 & 1e-2 & 10 & 1 & 0.05 & 1& RDMHD - no RSOL ($\tilde{c}=c$)\\
n1e22\_eps1\_rhd\_hr & & 1 & 1e-2 & 10 & 1 & 0.05 & 1 & RDMHD - higher spatial resolution\\
n1e22\_eps1\_rhd\_modB &  & 1 & 6.5e-3 & 27.0 & 5.7 & 0.11 & 0.13& RDMHD - weaker $\bf B^0$\\
n1e22\_eps1\_rhd\_modB\_lr &  & 1 & 6.5e-3 & 27.0 & 5.7 & 0.11 & 0.13& RDMHD - weaker $\bf B^0$ \& lower spatial resolution\\
\hline
n1e24\_eps1 & 1e+24 & 1 & 6.5e-5 & 2.7 & 5.7 & 0.11 & 0.13& default\\
n1e24\_eps0.1 & & 0.1 & 6.5e-6 & 27.0 & 5.7 & 1.1 & 0.13& 10x smaller $\epsilon_\text{grain}^{\text{max}}$\\ 
n1e24\_eps0.01 &  & 0.01 & 6.5e-7 & 270.0 & 5.7 & 11.0 & 0.13& 100x smaller $\epsilon_\text{grain}^{\text{max}}$\\
n1e24\_eps0.01\_hr\_t & & 0.01 & 6.5e-7 & 270.0 & 5.7 & 11.0 & 0.13& higher temporal resolution\\
n1e24\_eps0.01\_lr & & 0.01 & 6.5e-7 & 270.0 & 5.7 & 11.0 & 0.13& lower spatial resolution\\
n1e24\_eps0.01\_xlr & & 0.01 & 6.5e-7 & 270.0 & 5.7 & 11.0 & 0.13& much lower spatial resolution\\

\hdashline
n1e24\_eps1\_rhd & & 1 & 1e-4 & 1 & 1 & 0.05 & 1& default RDMHD \\
n1e24\_eps1\_rhd\_c & & 1 & 1e-4 & 1 & 1 & 0.05 & 1& No RSOL ($\tilde{c}=c$)\\
n1e24\_eps1\_rhd\_ fw & & 1 & 6.5e-4 & 2.7 & 5.7 & 0.02 & 1& lower $F_{\rm rad}$ \& stronger $\bf g$\\
\hline
n1e25\_eps1 & 1e+25& 1 & 6.5e-6 & 0.85 & 5.7 & 0.11 & 0.13& default\\ n1e25\_eps0.1 & & 0.1 & 6.5e-7 & 8.5 & 5.7 & 1.1 & 0.13& 10x smaller $\epsilon_\text{grain}^{\text{max}}$\\ 
n1e25\_eps0.01 &  & 0.01 & 6.5e-8 & 85.0 & 5.7 & 11.0 &0.13 & 100x smaller $\epsilon_\text{grain}^{\text{max}}$\\ 
\hdashline
n1e25\_eps1\_rhd & & 1 & 1e-5 & 0.32 & 1 & 0.05 & 1& default RDMHD\\
n1e25\_eps1\_rhd\_ fw & & 1 & 6.5e-6 & 0.85 & 5.7 & 0.02 & 0.13 & lower $F_{\rm rad}$ \& stronger $\bf g$\\

\hline
n1e26\_eps1 & 1e+26& 1 & 6.5e-7 & 0.27 & 5.7 & 0.11 & 0.13& default\\ 
n1e26\_eps0.1 & & 0.1 & 6.5e-8 & 2.7 & 5.7 & 1.1 & 0.13 & 10x smaller $\epsilon_\text{grain}^{\text{max}}$\\ 
n1e26\_eps0.01 & & 0.01 & 6.5e-9 & 27.0 & 5.7 & 11.0 & 0.13& 100x smaller $\epsilon_\text{grain}^{\text{max}}$\\ 

\hdashline
n1e26\_eps1\_rhd\_c & & 1 & 1e-6 & 0.1 & 1 & 0.05 & 1&default RDMHD\\
n1e26\_eps1\_rhd\_hr & & 1 & 1e-6 & 0.1 & 1 & 0.05 & 1 & RDMHD - higher spatial resolution \\
n1e26\_eps1\_rhd & & 1 & 1e-6 & 0.1 & 1 & 0.05 & 1& RDMHD - higher spatial resolution \& RSOL\\ 
n1e26\_eps1\_rhd\_modB & & 1 & 6.5e-7 & 0.27 & 5.7 & 0.11 & 0.13 & weaker $\bf B^0$\\ 
\hline
\end{tabular}
\caption{Initial conditions for all simulations. The simulations organized by the gas column density, and dashed lines separate simulations using the uniform flux approximation from the RDMHD runs. 
Columns show: 
{\bf (1)} Simulation name. 
{\bf (2)} Physical column density of the gas: $N_\text{gas}$.
{\bf (3)} Physical size of the largest grains: $\grainsize$.
{\bf (4)} Grain charge parameter $\chargeparammax \equiv 3\,\initvalupper{ \grainchargeZ }[\grainsizemax] \,e / (4\pi\,c\,(\grainsizemax)^{2}\,\rhobase^{1/2})$ of the largest grains. 
{\bf (5)} grain size parameter $\sizeparammax \equiv (\internaldensity\,\grainsizemax) / (\rhobase\,\Lscale)$ of the largest grains.
{\bf (6)} Gravity parameter $\gravparam \equiv |{\bf g}|\,\Lscale/\cs^{2}$.
{\bf (7)} Dust acceleration parameter: $\accparam \equiv (3/4)\,(F_{0} \langle Q \rangle_{\rm ext}\, /c )/(\rhobase\,\cs^{2})$. {\bf (8)} Initial plasma $\beta_{0} \equiv (\cs/\vA^{0})^{2}$. {\bf (9)} Notes for each run and key differences relative to its corresponding default run.}
\label{table:sims}
\end{table*}
\end{center}
\end{appendix}
\end{document}